\documentclass[12pt]{article}
\usepackage{amsmath}
\usepackage{amssymb}
\usepackage{subfigure}
\usepackage{latexsym}
\usepackage{hyperref}
\usepackage[sort]{cite}

\def\ket#1{\left|#1\right>}
{\catcode`\|=\active 
  \gdef\Braket#1{\left<\mathcode`\|"8000\let|\bravert {#1}\right>}}
\def\bravert{\egroup\,\vrule\,\bgroup}
\newcommand{\alg}[1]{\mathfrak{#1}}

\newcommand{\su}{\alg{su}}

\newcommand{\Sl}{\alg{sl}}
\newcommand{\so}{\alg{so}}

\newcommand{\be}{\begin{eqnarray}}
\newcommand{\ee}{\end{eqnarray}}
\newcommand{\bc}{\begin{center}}
\newcommand{\ec}{\end{center}}
\newcommand{\bea}{\begin{eqnarray}}
\newcommand{\eea}{\end{eqnarray}}
\newcommand{\ben}{\begin{equation}}
\newcommand{\een}{\end{equation}}

\newcommand{\del}{\partial}

\newcommand{\nn}{\nonumber}

\newcommand{\la}{\label}
\newcommand{\vc}{\bar v}

\newcommand{\Mgamma}{\gamma}
\newcommand{\METgamma}{{\tilde\gamma}}
\def\pint{{-\!\!\!\!\!\!\int}}
\numberwithin{equation}{section}

\begin{document}

\begin{titlepage}
\begin{flushright}
hep-th/0605018
\end{flushright}
\vspace{8 mm}
\begin{center}
{\Large \bf  Integrable twists in AdS/CFT }
\end{center}
\vspace{6 mm}

\begin{center}
{ Tristan McLoughlin${}^a$ and Ian Swanson${}^b$ }\\
\vspace{6mm}
${}^a${\it Department of Physics, Pennsylvania State University \\
University Park, PA 16802, USA }\\
\vspace{6mm}
${}^b${\it School of Natural Sciences, Institute for Advanced Study\\
Princeton, NJ 08540, USA }
\end{center}
\vspace{6 mm}
\begin{center}
{\large \bf Abstract}
\end{center}
\noindent
{\small
A class of marginal deformations of four-dimensional 
${\cal N}=4$ super Yang-Mills theory has been found
to correspond to a set of smooth, multiparameter deformations of the 
$S^5$ target subspace in the holographic dual on $AdS_5\times S^5$.  
We present here an analogous set of deformations that act on global toroidal 
isometries in the $AdS_5$ subspace. 
Remarkably, certain sectors of the string theory remain classically integrable
in this larger class of so-called $\gamma$-deformed $AdS_5\times S^5$ backgrounds.
Relying on studies of deformed $\su(2)_\gamma$ models, 
we formulate a local $\Sl(2)_\gamma$ Lax representation that admits a classical, thermodynamic 
Bethe equation (based on the Riemann-Hilbert interpretation of Bethe's ansatz) 
encoding the spectrum in the deformed $AdS_5$ geometry.  
This result is extended to a set of discretized, asymptotic Bethe equations
for the twisted string theory.  Near-pp-wave energy spectra within
$\Sl(2)_\gamma$ and $\su(2)_\gamma$ sectors
provide a useful and stringent test of such equations, 
demonstrating the reliability of this technology in a wider class of string 
backgrounds.  In addition, we study a twisted Hubbard model that yields
certain predictions of the dual $\beta$-deformed gauge theory.}
\end{titlepage}
\tableofcontents
\newpage
\section{Introduction}
In \cite{hep-th/0502086}, Lunin and Maldacena used an $SL(3,R)$ deformation of $AdS_5\times S^5$ to
find a supergravity solution dual to a class of marginal deformations 
(known as Leigh-Strassler \cite{hep-th/9503121} or $\beta$-deformations) of ${\cal N}=4$
super Yang-Mills (SYM) theory.  This provided an interesting opportunity to study the AdS/CFT 
correspondence \cite{hep-th/9711200,hep-th/9802109,hep-th/9802150} in new gravity backgrounds with less
supersymmetry.  In the case of real deformations, one obtains the gravity dual of a 
one-parameter family of ${\cal N}=1$ conformal gauge theories, and this particular example has been 
the focus of many recent investigations: pp-wave limits were studied in 
\cite{hep-th/0505227,hep-th/0505243}, for example, and other interesting string systems 
were examined in 
\cite{hep-th/0509036,hep-th/0503192,hep-th/0506063,hep-th/0507136,hep-th/0509195,hep-th/0509058,hep-th/0511216}. 

The notion of the Lunin-Maldacena deformation was generalized by Frolov 
\cite{hep-th/0503201} by considering a sequence of T-dualities and coordinate
shifts, or TsT deformations, acting on global toroidal isometries in $S^5$.
By parameterizing each TsT deformation with separate $\METgamma_i$ $(i \in 1,2,3)$, 
one generically obtains a non-supersymmetric theory, dual to  
a non-supersymmetric deformation of ${\cal N}=4$ SYM.\footnote{Of course, 
the $\METgamma_i$ can be chosen to
reproduce the Lunin-Maldacena solution as a special case.}
(Adhering to conventions in the literature, we will use the symbols $\METgamma_i$
to indicate deformation parameters that naturally appear in the background geometry.)
This construction can be extended to include complex deformations by including 
$SL(2,R)$ transformations.  
By studying string theory on $AdS_5$ backgrounds with TsT-deformed 
$S^5$ factors, Frolov was also able to demonstrate that bosonic string solutions in these
backgrounds can be generated by imposing twisted boundary conditions on known
solutions in the undeformed $AdS_5\times S^5$ geometry. 
The full action for Green-Schwarz strings in TsT-deformed backgrounds 
was subsequently constructed in \cite{hep-th/0512253}, where it was shown that
{\it superstring} solutions in such backgrounds are again mapped (in a one-to-one fashion) 
from solutions in the parent geometry, deformed by twisted boundary conditions.

Another interesting property of TsT transformations in $S^5$ is that the integrability of classical
string theory on $AdS_5\times S^5$ \cite{hep-th/0206103,hep-th/0305116} seems to be preserved under
these deformations.  In \cite{hep-th/0503192}, Frolov, Roiban and Tseytlin were able to derive 
classical Bethe equations encoding the spectral problem in (classically) closed sectors 
on the deformed $S^5$ subspace.   Similar to the undeformed case 
\cite{hep-th/0402207,hep-th/0410105,hep-th/0502226,hep-th/0410253,hep-th/0412254,hep-th/0502240}, 
Frolov \cite{hep-th/0503201} was subsequently able to derive a Lax representation
for the bosonic sector of the deformed-$S^5$ theory:
the essential observation was that one can gauge away the non-derivative 
dependence of the Lax representation on the $U(1)$ isometry fields involved in the deformation.

In addition to deriving twisted Bethe equations,  Frolov, Roiban and Tseytlin demonstrated
in \cite{hep-th/0503192} that more general 
fast-string limits in these deformed backgrounds can be described by a 
Landau-Lifshitz action corresponding to a continuum limit 
of anisotropic spin chains associated with the scalar sector of the deformed 
${\cal N}=1$ dual gauge theory\footnote{Notions of integrability on the gauge theory
side of the correspondence are understood here to be restricted to the planar limit.} 
\cite{hep-th/0312218,hep-th/0405215}.
Various aspects of these twisted spins chains have been studied in 
\cite{hep-th/0507136,hep-th/0510221,hep-th/0511164,hep-th/0601109}, for example.
In this vein, Beisert and Roiban provided a detailed study of 
related spin-chain systems with a variety of twists in \cite{hep-th/0505187} that will be
particularly useful in the present context.

In this paper we add to previous studies of semiclassical strings in $\gamma$-deformed backgrounds.
We focus largely on deformations of the $AdS_5$ subspace analogous to 
Frolov's multiparameter TsT deformations.  
We find a one-parameter family of such deformations that can
be understood as a usual TsT deformation acting on a global $U(1)\times U(1)$ isometry in
$AdS_5$, while a wider class of deformations  
can be seen as arising from TsT transformations that 
involve T-duality along timelike directions.  As with the $S^5$ deformations, 
however, we can again interpret such transformations as formally giving rise to 
twisted boundary conditions from the perspective of the undeformed theory, and
integrability in the classical string theory is again preserved.  
In these new geometries we find a number of peculiar features, and 
we expect the dual gauge theory to be modified dramatically (perhaps to 
a non-commutative gauge theory, along the lines
of \cite{hep-th/9907166,hep-th/9908134}).  

In section two we establish notation, review TsT deformations on the $S^5$ subspace and 
parameterize the geometry in a way that is convenient for
studying a pp-wave limit of the deformed background.  In section three we present 
analogous deformations of the $AdS_5$ subspace and study properties of the
resulting geometry.  We study a suitable Lax representation for string 
theory on this background in section four, and compute the Riemann-Hilbert formulation of the 
classical Bethe equations within deformed $\Sl(2)_\gamma$ sectors of the theory.  Energy
spectra in the near-pp-wave limit of $\gamma$-deformed $\su(2)_\gamma$ and $\Sl(2)_\gamma$ sectors 
are computed in section five.  In section six we study $\gamma$-deformed, discrete 
extrapolations of the thermodynamic string Bethe equations, and comment
on the ability of these equations to reproduce the near-pp-wave energy spectra of 
BMN strings in these backgrounds.  In section seven we study a twisted Hubbard model, 
analogous to \cite{hep-th/0512077}, that is conjectured to yield the deformed $\su(2)_\gamma$ sector
of the dual gauge theory.  In the case of the corresponding deformed $\Sl(2)_\gamma$ sector,
we have little to say regarding predictions from the gauge theory side of the
correspondence.  In the final section, however, we comment on various relevant 
Bethe equations proposed in \cite{hep-th/0505187}.

\section{Geometry and TsT deformations}
\label{su2DEF}
TsT transformations correspond to a sequence of worldsheet duality transformations,
and one expects deformed backgrounds obtained in this manner 
to be exact solutions of the equations of motion.   By including S-duality 
one can extend this class of transformations (parameterized by real $\METgamma_i$)
to include complex deformation parameters. (Backgrounds obtained in this fashion, however, 
are expected to be modified by $\alpha'/R^2$ corrections \cite{hep-th/0503192}.)  
For the case of real $\METgamma_i$, the deformed spacetime metric 
and relevant background fields are given by (mostly following the notation of 
\cite{hep-th/0507021}), 
\begin{eqnarray}
ds^2_{\rm string}/R^2 &=& ds^2_{{AdS_5}} +
   \sum^3_{i=1} ( d\rho_i^2  + G \rho_i^2 d\phi_i^2) +   G
\rho_1^2\rho_2^2\rho_3^2 [d (\sum^3_{i=1} \METgamma_i \phi_i)]^2\ ,
\label{me}
\nn\\
\la{bfield0}
B_2 &=&  R^2  G w_2 \ , 
	\qquad 
	e^\phi = e^{\phi_0}G^{1/2}\ , 
	\qquad \chi = 0\ ,
\nn\\
w_2  &\equiv&  \METgamma_3\,  \rho_1^2 \rho_2^2\, d\phi_1 \wedge d\phi_2 +
\METgamma_1\,  \rho_2^2 \rho_3^2\, d\phi_2 \wedge d\phi_3 + \METgamma_2\,
\rho_3^2 \rho_1^2\, d\phi_3 \wedge d\phi_1\ ,  \nn\\
\label{ofggen0}
G^{-1} &\equiv& 1 +  \METgamma_3^2\,  \rho_1^2 \rho_2^2 +  \METgamma_1^2\,
\rho_2^2 \rho_3^2 +  \METgamma_2^2\,  \rho_1^2 \rho_3^2\ ,
\end{eqnarray}
where we have set $\alpha'=1$ for convenience.
The component $ds^2_{{AdS_5}}$ represents the undeformed metric on the
$AdS_5$ subspace, while the $S^5$ subspace has undergone three consecutive
TsT deformations parameterized by the $\METgamma_i$.  
To be certain, the $\METgamma_i$ appearing in the metric parameterize the coordinate-shift 
part of individual TsT deformations 
($\phi_1 \to \phi_1 + \METgamma_i\, \phi_2$, for example). 
$B_2$ is the NS-NS two-form field
strength (we have omitted the two- and five-form field strengths $C_2$ and 
$F_5$).  The usual angle variables on the sphere can be encoded in the 
convenient parameterization
\be
\label{angl}
\rho_1 = \sin \alpha \cos \theta\ , \ \ \
\rho_2 =  \sin \alpha \sin \theta\ , \ \ \
 \rho_3=  \cos \alpha\ ,
\ee
such that $\sum_{i=1}^3\rho_i^2=1$.
The string coupling $g_s$ is related to the gauge theory coupling 
$g_{\rm YM}$ via the standard relation, 
$g_s =  e^{\phi_0}  =  {g^2_{\rm YM}/ 4\pi}$, and the radial scale of 
both the $AdS_5$ and $S^5$ spaces is given by
$R^4= 4 \pi g_s N_c = g^2_{\rm YM}N_c \equiv\lambda$, where $N_c$ is the rank of 
the Yang-Mills gauge group.  We will restrict our attention to the full
planar limit $N_c \to \infty$. 

For present purposes, we find it convenient to introduce the 
following alternative parameterization on $S^5$:
\be
\label{ppparam}
    \rho_2  = \frac{y_1}{R}~, \qquad
    \rho_3  =  \frac{y_2}{R}\ , \qquad
    \rho_3  = \sqrt{1-\rho_2^2-\rho_3^2}\ , \qquad
    \phi_1  = x^+ +\frac{x^-}{R^2}\ , \qquad
    t = x^+\ .
\ee
This choice of lightcone coordinates implies that, as $R$ becomes large,
we approach a semiclassical limit described by point-like 
(or ``BMN'' \cite{hep-th/0202021}) strings boosted to lightlike momentum 
$J$ along a geodesic on the deformed $S^5$.
The angular momentum $J$ in 
the $\phi_1$ direction is related to the scale radius $R$ according to  
\be
p_- R^2 = J\ ,
\ee
and the lightcone momenta take the form
\be
-p_+ = \Delta-J~, \qquad -p_-=i\partial_{x^-} = \frac{i}{R^2}\partial_\phi 
	= -\frac{J}{R^2}\ .
\ee

At this stage we find it convenient to use the following form of the $AdS_5$ metric:
\be 
\la{adsmetric0}
ds^2_{AdS_5}=-\left(\frac{1+ x^2/4 R^2}{1- x^2/4 R^2}\right)^2 dt^2 
	+ \frac{dx^2/R^2}{(1-x^2/4 R^2)^2}\ .
\ee
This version of the spacetime metric is also useful when working with fermions
(see \cite{hep-th/0307032,hep-th/0404007} for details), though
we will restrict ourselves to the bosonic sector of the string theory in the
present study.  The coordinates $x^k$, $y_1^{k'}$ and $y_2^{k'}$ span
an $SO(4)\times SO(2)\times SO(2)$ transverse space, with $z_k$ lying in $AdS_5$ 
($k \in 1,\ldots ,4$), and $y_1^{k'},\ y_2^{k'}$ parameterizing the deformed 
$S^5$ subspace ($k' \in 1,2$).
In the pp-wave limit, $p_-$ is held fixed while $J$ and $R$ become infinite,
and the planar limit is taken such that the quantity $N_c/J^2$ is held fixed.
The lightcone momentum $p_-$ is then equated with
\be
p_- = \frac{1}{\sqrt{\lambda'}} = \frac{J}{\sqrt{g_{\rm YM}^2 N_c}}\ ,
\ee
where $\lambda'$ is known as the modified 't~Hooft coupling, 
and $J$ is equated on the gauge theory side with a scalar component of the
$SU(4)$ $R$-charge.

For reasons described in \cite{hep-th/0307032,hep-th/0404007}, 
the lightcone coordinates in eqn.~(\ref{ppparam}) admit
many simplifications, the most important of which is
the elimination of normal-ordering contributions to the 
lightcone Hamiltonian.  (More recently, however, 
a ``uniform'' lightcone gauge choice has proved
to be useful in certain contexts \cite{hep-th/0510208}.) 
It is convenient to introduce the following complex coordinates
\be
\label{complex}
y&=& y_1 \cos \phi_2+i y_1 \sin \phi_1\ , \qquad  
	\bar{y}=y_1 \cos \phi_2 - i y_1 \sin \phi_1\ ,\nn\\
z&=& y_2 \cos \phi_2+i y_2 \sin \phi_1\ , \qquad  
	\bar{z}=y_2 \cos \phi_2 - i y_2 \sin \phi_1\ .
\ee
Arranging the large-$R$ expansion of the spacetime metric according to
\be
ds^2 = ds_{(0)}^2 + \frac{ds_{(1)}^2}{R^2} + O(1/R^4)\ ,
\ee
we therefore find
\be
\label{expndmet0}
ds_{(0)}^2&=&2dx^+dx^- + |dy|^2 + |dz|^2 -(dx^+)^2
	\left[x^2+|y|^2(1+\METgamma_3^2)+|z|^2(1+\METgamma^2_2)\right]\ ,
\nn\\
ds_{(1)}^2 &=& 
	(dx^-)^2+\frac{1}{4}(yd\bar{y}+\bar{y}dy+zd\bar{z}+\bar{z}dz)^2
	-2dx^+ dx^-(|y|^2(1+\METgamma_3^2)+|z|^2(1+\METgamma_2^2))  
\nn\\
& & +\frac{1}{2}x^2dx^2 
	+(dx^+)^2\Bigl[(-\frac{1}{2}x^4 +2(|z|^2+|y|^2)(|z|^2\METgamma_2^2
	+|y|^2\METgamma_3^2)+(|y|^2\METgamma_3^2+|z|^2\METgamma_2^2)^2\Bigr]
\nn\\
& &+\METgamma_1 dx^+(\METgamma_2 |z|^2 \Im(\bar{y}dy)+\METgamma_3 |y|^2 
	\Im(\bar{z}dz))-\METgamma_3^2\Im(\bar{y}dy)^2
	-\METgamma_3^2\Im(\bar{z}dz)^2
\nn\\
& &+ \METgamma_2\METgamma_3\Im(\bar{y}dy)\Im(\bar{z}dz)\ .
\ee 
At leading order one obtains a pp-wave metric, with obvious contributions
from the deformed and undeformed subspaces.  At both leading and sub-leading order
near the pp-wave limit, we find that the geometry is deformed only by the parameters
$\METgamma_2$ and $\METgamma_3$ (this is just a consequence of the 
particular semiclassical limit we have chosen).
The corresponding expansion of the NS-NS two-form $B_2$ appears as
\be
B_2&=&\METgamma_3 dx^+\wedge \Im (\bar{y}dy)-\METgamma_2 dx^+ \wedge \Im(\bar{z}dz)\nn\\
& &+\frac{1}{R^2}\left[ -\METgamma_3(\METgamma_3^2|y|^2+\METgamma_2^2|z|^2)dx^+\wedge
\Im(\bar{y}dy)+\METgamma_2(\METgamma_3^2|y|^2
	+\METgamma_2^2|z|^2)dx^+\wedge\Im(\bar{z}dz)\right.\nn\\
& & \left. +\METgamma_3 dx^-
  \Im(\bar{y}dy)-\METgamma_2dx^-\wedge\Im(\bar{z}dz)+\METgamma_1\Im(\bar{y}dy)\wedge\Im(\bar{z}dz)\right]\ .
\ee

We truncate to $\su(2)_\gamma$ sectors of the geometry by projecting onto
a single complex coordinate, which isolates a one-parameter TsT deformation:
\be
ds^2_{\su(2)_\gamma}&=&2 dx^+dx^--\left(1+\METgamma^2\right)|y|^2 (dx^+)^2+|dy|^2+\frac{1}{R^2}
	\Big[ \frac{1}{4}(yd{\bar y}
	+{\bar y}dy)^2
\nn\\ 
&&\kern-20pt
	+(dx^-)^2+\METgamma^2(2+\METgamma^2)|y|^4(dx^+)^2-2(1+\METgamma^2)|y|^2dx^+dx^-
        -\METgamma \Im\left({\bar y}dy\right)\Big] + O(1/R^4)\ .
\nn\\
&&
\ee
The parameter $\METgamma$ here can stand for either $\METgamma_2$ or $\METgamma_3$, 
corresponding to two possible choices of $\su(2)_\gamma$ truncation.  
The NS-NS two-form reduces in this $\su(2)_\gamma$ sector to
\be
B_2^{\su(2)_\gamma}&=&\METgamma dx^+ \wedge \Im({\bar y}dy)+\frac{\METgamma}{R^2}\Bigl(dx^-\wedge \Im({\bar y}dy)
\nn\\
&&
	-(1+\METgamma^2)|y|^2 dx^+ \wedge \Im({\bar y}dy)\Bigr)+ O(1/R^4)\ .
\ee
Given this parameterization we can easily calculate the string lightcone Hamiltonian and
solve for its (semiclassical) spectrum.  Prior to attacking this problem, we will formulate an 
analogous deformation on the $AdS_5$ subspace.

\section{Deformations of $AdS_5$}
\label{AdSDEF}
To study TsT deformations on $AdS_5$, we find it convenient to parameterize
the geometry in a nearly identical fashion to the $S^5$ case above, such that the spacetime
metric in each individual subspace is mapped into the other under an obvious Wick rotation.
It is useful in this regard to start with an $SO(4,2)$ invariant, 
expressed in terms of ${\mathbb R}^6$ embedding coordinates,
\be
\label{so42inv}
-X_0^2 + X_1^2 + X_2^2+ X_3^2+ X_4^2- X_5^2 = -1\ ,
\ee
which we write as
\be
X_0 = \eta_1 \sin\hat\varphi_1\ , &\qquad & X_1 = \eta_2\cos\hat\varphi_2\ , \nn\\
X_2 = \eta_2 \sin\hat\varphi_2\ , &\qquad & X_3 = \eta_3\cos\hat\varphi_3\ , \nn\\
X_4 = \eta_3 \sin\hat\varphi_3\ , &\qquad & X_5 = \eta_1\cos\hat\varphi_1\ . 
\ee
The hatted notation $\hat\varphi_i$ is employed to denote untwisted 
$U(1)$ angular coordinates.  This formulation is connected with the usual angular 
variables on $AdS_5$ under the assignment
\be
\eta_1 = \cosh\alpha\ , 
	\qquad 
	\eta_2 = \sinh\alpha \sin\theta\ ,
	\qquad 
	\eta_3 = \sinh\alpha\cos\theta\ ,
\ee
which preserves the $SO(2,1)$ invariant
\be
\label{so21inv}
- \eta_1^2 + \eta_2^2 + \eta_3^2 = -1\ .
\ee
One thereby obtains the more familiar spacetime metric:
\be
{ds_{AdS_5}^2}/{R^2}
	 & = &
	-(d\eta_1^2 + \eta_1^2 d\hat\varphi_1^2)
	+ \sum_{i=2}^3 (d\eta_i^2 + \eta_i^2 d\hat\varphi_i^2) 
\nn\\
& = & 	 d\alpha^2 - \cosh\alpha^2 d\hat\varphi_1^2
	+ \sinh\alpha^2 \left(
	d\theta^2 + \sin\theta^2 d\hat\varphi_2^2 
	+ \cos\theta^2d\hat\varphi_3^2 \right)\ .
\ee
This metric exhibits a manifest $U(1)\times U(1)\times U(1)$ global symmetry:
the deformations of interest thus act on the corresponding 
angular coordinates $\hat\varphi_i$ ($i\in 1,2,3$).  At this stage one may be concerned that 
invoking T-duality on compact timelike directions will lead to complications.
While this concern will be addressed below, we aim to simplify the discussion 
by focusing on a single TsT deformation for which this issue can be avoided 
under certain assumptions.
Following any manipulations, of course, one must pass to the universal 
covering space in which the global time coordinate in the resulting geometry is 
understood to be noncompact.

\subsection{Single-parameter TsT deformation }
It turns out that TsT transformations on the angular pair
$(\hat\varphi_2,\hat\varphi_3)$ result in a deformation that is trivial upon 
reduction to deformed $\Sl(2)_\gamma$ subsectors of the theory.    
The global $U(1)\times U(1)$ isometry of interest is thus chosen to be 
parameterized by the angular coordinates $\hat\varphi_1$ and $\hat\varphi_2$.
The $\hat\varphi_2$ direction is spacelike, so we invoke a TsT transformation 
that acts as a T-duality along the $\hat\varphi_2$ direction, a shift in the $\hat\varphi_1$
direction $\hat\varphi_1 \to \hat\varphi_1 + \METgamma \hat\varphi_2$, followed by a second T-duality 
in the new $\varphi_2$ direction (where $\varphi_2$ now stands for a ``transformed'' angular 
coordinate).   The deformed spacetime metric thus takes the form
\be
\label{DEFMET}
ds^2_{\rm str}/R^2 =  ds^2_{_{S^5}} +
      g^{ij} d\eta_i d\eta_j  
		+ g^{ij}\, G\, \eta_i^2 d\varphi_j^2  
	- \METgamma^2\, G\,  \eta_1^2\eta_2^2\eta_3^2\,  d  \varphi_1^2 \ ,
\ee
where $g = {\rm diag}(-1,1,1)$, and the deformation factor $G$ is given by
\be
G^{-1} &\equiv& 1 - \METgamma^2  \eta_1^2 \eta_2^2\ .
\ee
The NS-NS two-form appears as
\be
B_2/R^2 = \METgamma\, G\, \eta_1^2 \eta_2^2\, d\varphi_1 \wedge d\varphi_2\ .
\ee
An obvious concern is that one has generated spacetime directions in this
background that exhibit mixed signature.  In fact,  
immediately following the shift in the $\hat\phi_1$ direction, the dualized
$\phi_2$ direction becomes mixed.  
From the worldsheet perspective, the string action should only be 
sensitive to the local region of the target space in which the string propagates,
so we may choose to study the theory in the region 
where $\phi_2$ is strictly spacelike. In any case, we find it efficient (for the moment)
to proceed pragmatically by studying the theory in regions of the geometry where the 
signature is unambiguous.  It will be shown below that this naive approach yields a 
lightcone Hamiltonian that appears to be sensible for our purposes.

Under the deformation in eqn.~({\ref{DEFMET}), 
the $AdS_5^\METgamma$ worldsheet action can be written as
\be
S_{AdS_5^\METgamma} & = & -\frac{\sqrt{\lambda}}{2}\int d\tau \frac{d\sigma}{2\pi}
	\Bigl[ \gamma^{\alpha\beta}
	\left(
	g^{ij} \del_\alpha\eta_i \del_\beta\eta_j  
		+ g^{ij}\, G\,  \eta_i^2 \del_\alpha \varphi_j \del_\beta\varphi_j  
	- \METgamma^2\, G\,  \eta_1^2\eta_2^2\eta_3^2\,  
	\del_\alpha\varphi_1\del_\beta\varphi_1
	\right)  
\nn\\
&&
\kern+50pt
	-2 \epsilon^{\alpha\beta}\left(
	\METgamma\, G\, \eta_1^2\eta_2^2\, \del_\alpha \varphi_1 \del_\beta \varphi_2
	+ \Lambda (g^{ij}\eta_i\eta_j +1 )
	\right)\Bigr]\ ,
\ee
where $\Lambda$ acts as a Lagrange multiplier enforcing eqn.~(\ref{so21inv}) 
on shell.  The indices $\alpha$ and $\beta$ run over the $\tau$ ($\alpha,\ \beta = 0$) and $\sigma$ 
($\alpha,\ \beta = 1$) directions
on the worldsheet, and $\gamma^{\alpha\beta}$ is the worldsheet metric.  
With the intent of studying semiclassical limits of this action, it is  
convenient to choose a lightcone coordinate parameterization analogous 
to eqn.~(\ref{ppparam}) above:
\be
    \eta_2  = \frac{u_1}{R}~, \qquad
    \eta_3  =  \frac{u_2}{R}\ , \qquad
    \eta_3  = \sqrt{1+\eta_2^2+\eta_3^2}\ , \qquad
    \phi_1  = x^+ +\frac{x^-}{R^2}\ , \qquad
    t = x^+\ ,
\ee
and rewrite the metric on $S^5$ (with an $SO(4)$ coordinate $s$) as
\be
ds^2_{S^5}=\left(\frac{1 - s^2/4 R^2}{1+ s^2/4 R^2}\right)^2 dt^2 
+ \frac{ds^2/R^2}{(1-s^2/4 R^2)^2}\ .
\ee
By analogy with eqn.~(\ref{complex}), we introduce the following complex
coordinates
\be
\label{complex2}
v&=& u_1 \cos \varphi_2+i u_1 \sin \varphi_1\ , \qquad  \bar{v}=u_1 \cos \varphi_2 
	+ i u_1 \sin \varphi_1\ ,\nn\\
w&=& u_2 \cos \varphi_2+i u_2 \sin \varphi_1\ , \qquad  \bar{w}=u_2 \cos \varphi_2 
	+ i u_2 \sin \varphi_1\ ,
\ee
so that truncation to deformed $\Sl(2)_\gamma$ sectors involves projecting onto the
coordinate pairs $(v,\bar v)$ or $(w, \bar w)$.
In the $(v,\bar v)$ projection the large-$R$ expansion of the spacetime metric and 
NS-NS two-form yields:
\be
ds^2_{\Sl(2)_\gamma} & = & 2 dx^+ dx^- -  (1+\METgamma^2)|v|^2 (dx^+)^2 + |dv|^2
	- \frac{1}{R^2}\Bigl[
	\frac{1}{4}(v d{\bar v}
	+{\bar v}dv)^2 - (dx^-)^2 
\nn\\
&&	+ \METgamma^2(2+\METgamma^2)|v|^4(dx^+)^2
        + \METgamma^2 (v d\bar v - \bar v dv)^2 \Bigr] + O(1/R^4)\ ,
\nn\\
B_2^{\Sl(2)_\gamma} &=& \frac{i}{2}\METgamma dx^+ \wedge (v d\bar v - \bar v dv)
	+\frac{i}{2R^2} |v|^2 \METgamma (1+\METgamma^2) dx^+ \wedge
	(\bar v dv - v d\bar v)+ O(1/R^4)\ .
\nn\\
&&
\ee

As with the corresponding deformations on $S^5$, the TsT deformation considered here 
amounts to a set of twisted boundary conditions on the undeformed $U(1)$ coordinates.  As in 
\cite{hep-th/0503201}, one finds that the conserved $U(1)$ currents 
$J_i^\alpha$ in the undeformed theory are identical to those in the deformed theory.  
Labeling canonical momenta as $p_i = J_i^0$, the associated charges take the form
\be
J_i = \int \frac{d\sigma}{2\pi} p_i\ .
\ee
One therefore finds that 
the identification of the deformed and undeformed currents $J_i^1$ leads to the conditions
\be
\label{TBC0}
\hat\varphi_1' & =  &  \varphi_1' - \Mgamma\, p_2\ , \nn\\
\hat\varphi_2' & =  &  \varphi_2' + \Mgamma\, p_1\ , \nn\\
\hat\varphi_3' & =  &  \varphi_3'\ ,
\ee
where $\varphi'$ denotes a worldsheet $\sigma $ derivative acting on $\varphi$,
and, for convenience (and to remain consistent with the literature), 
we have introduced the rescaled deformation parameter
\be
\label{gammascale}
\Mgamma \equiv \frac{\METgamma}{\sqrt{\lambda}}\ .
\ee
In the case of $S^5$ deformations, this quantity is to be identified with $\beta$, 
which is the deformation parameter in the corresponding $\beta$-deformed gauge theory.

These equations imply the
following twisted boundary conditions on the undeformed $U(1)$ coordinates:
\be
\label{TBC}
\hat\varphi_1(2\pi) - \hat\varphi_1(0) & = & 2\pi(m_1 -\Mgamma\, J_2)\ , \nn\\
\hat\varphi_2(2\pi) - \hat\varphi_2(0) & = & 2\pi(m_2 +\Mgamma\, J_1)\ , \nn\\
\hat\varphi_3(2\pi) - \hat\varphi_3(0) & = & 2\pi m_3\ , 
\ee
where $m_i$ here denotes integer winding numbers defined by
\be
2\pi m_i = \varphi_i(2\pi) - \varphi_i(0)\ .
\ee
In other words, the TsT deformation in eqn.~(\ref{DEFMET}) amounts to imposing the
above boundary conditions on the angular variables $\hat\varphi_i$, in precise 
analogy with corresponding deformations of $S^5$.

\subsection{Euclideanized deformation}
It was suggested in \cite{hep-th/0505187} that the above deformation of $AdS_5$
might correspond to a non-commutative deformation of the dual Yang-Mills theory.
To explore this possibility, it is useful to consider TsT deformations in Euclidean $AdS_5$.
Starting from a Wick rotation of eqn.~(\ref{so42inv})
\be
-X_0^2+X_5^2+X_1^2+X_2^2+X_3^2+X_4^2=-1\ ,
\ee
with the parameterization
\be
X_0=\eta_1 \cosh \hat\varphi_1\ , &\qquad& X_5= \eta_1 \sinh \hat\varphi_1\ ,\nn\\
X_1=\eta_2 \cos \hat\varphi_2\ , &\qquad& X_2= \eta_2 \sin \hat\varphi_2\ ,\nn\\
X_3=\eta_3 \cos \hat\varphi_3\ , &\qquad& X_4= \eta_3 \sin \hat\varphi_3\ ,
\ee
one obtains the following undeformed Euclideanized spacetime metric:
\be
\frac{ds^2}{R^2}=-d\eta_1^2+\eta_1^2 d\hat\varphi_1^2+\sum_{i=2}^{3}(d\eta_i^2+\eta_i^2 
d\hat\varphi_i^2)\ .
\ee
A TsT transformation in the $(\hat\varphi_1,\hat\varphi_2)$ coordinates yields
\be
{ds^2}/{R^2}&=&-d \eta_1^2+G \eta_1^2d\varphi_1^2+\sum_{i=2}^3 (d \eta_i^2+G \eta_i^2d\varphi_i^2)
-G\ \eta_1^2\eta_2^2\eta_3^2\ \METgamma^2 d\varphi^2_3\ ,
\ee
with  $G^{-1}=1+\METgamma^2 \eta_1^2  \eta_2^2$, and  $B_{12}= \METgamma\ G\ \eta_1^2 \eta_2^2$.
Moving to the Poincar\'e coordinates 
\be
\eta_1=\frac{\sqrt{y^2+z_1^2+z_2^2}}{y}\ ,  \quad
\eta_2=\frac{z_1}{y}\ ,  \quad
\eta_3=\frac{z_2}{y}\ ,  \quad
\varphi_1 = {\rm{ln}} \sqrt{y^2+z_1^2+z_2^2}\ ,
\ee
we obtain the following metric:
\be
\label{EucMet}
ds^2/R^2 
	&=&\frac{1}{y^2}\left(dy^2+dz_1^2+dz_2^2\right)
	+\frac{G}{y^2}\left(z_1^2 
	d\varphi_2^2+z_2^2 d\varphi_3^2\right)
	-\frac{G \METgamma^2}{y^6}(y^2+z_1^2+z_2^2)
	z_1^2z_2^2d\varphi_3^2
\nn\\
&&
	- \left(\frac{\METgamma^2 z_1^2}{y^2(y^4+\METgamma^2 z_1^2(y^2+z_1^2+z_2^2))}\right)
	(ydy+z_1dz_1+z_2dz_2)^2\ ,
\ee
where 
\be
G=\frac{y^4}{y^4+\METgamma^2 z_1^2 (y^2+z_1^2+z_2^2)}\ .
\ee

Let us now consider various limits of this geometry.
As  $y \to \infty$, $G$ tends to unity, and one recovers the original undeformed geometry. 
This corresponds to an infrared limit in the dual
theory and, as one would expect, 
the physics should become insensitive in this limit 
to a non-commutativity parameter \cite{hep-th/9907166,hep-th/9908134}. 
In the $y\to 0$ limit $G$ scales according to
\be
G \sim \frac{y^4}{\METgamma^2 z_1^2 (z_1^2+z_2^2)}\ , \qquad (y \to 0)\ , 
\ee
so that one obtains
\be
\label{Euc_def_metric}
ds^2/R^2=
	\frac{1}{y^2}\left(dy^2+dz_1^2+dz_2^2\right)
	-\left(\frac{1}{y^2(z_1^2+z_2^2)}
	\right)(z_1dz_1+z_2dz_2)^2-\frac{1}{y^2}z_2^2d\varphi_3^2\ .
\ee
This corresponds to a UV limit, where  
we expect the dual theory to acquire modifications depending on the
non-commutativity parameter.

It is worthwhile  to note that the curvature invariants possess a certain scaling symmetry. 
In the $y\to 0$ limit, the metric in eqn.~(\ref{Euc_def_metric}) is symmetric under 
$y\to \lambda y$, $(z_1, z_2) \to (\lambda z_1, \lambda z_2)$. 
The curvature invariants therefore do not depend on $y$ as $y\rightarrow 0$, and thus remain 
bounded in this limit. 

While this geometry exhibits many of the properties known to persist in  
gravity duals of non-commutative Yang-Mills theories, 
there are some unexpected features as well. 
If we consider the subspace defined by $z_1=0$, we see that the 
$\METgamma$ dependence drops out of the metric entirely.
Furthermore, if we look at the scale where 
the deformation is noticeable, we find that $y\sim {1}/{\METgamma z_1}$.  The theory 
therefore appears to have a position-dependent non-commutativity scale.  Non-commutative gauge 
theories with a non-constant parameter have been studied in the context of the AdS/CFT
correspondence 
\cite{hep-th/0001203,hep-th/0005159,hep-th/0105191}, where they arise from non-vanishing NS-NS 
$H$ flux and give rise to a non-associative star product \cite{hep-th/0101219}. 
To be certain, it is useful to point out that we have allowed deformations to act on isometries 
corresponding to the Cartan generators of the four-dimensional conformal group.
This should be contrasted with the studies in \cite{hep-th/9907166,hep-th/9908134}, where deformations were
chosen to act on isometries on the boundary in a Poincar\'e patch \cite{hep-th/0505187}.
Although our focus here is primarily restricted to the stringy 
aspects of the correspondence, these considerations clearly present a number of interesting 
issues on the field theory side that remain to be explored.

\subsection{A note on finite temperature }
From the point of view of the deformed Euclidean $AdS_5$ metric
in eqn.~(\ref{EucMet}), 
it is not obvious what the finite-temperature extension should be. 
To study this question, we instead 
start from TsT deformations of the standard finite-temperature gravity solution. 

The non-extremal Euclideanized D3 brane background with temperature $T$ is given by
\be
ds^2/R^2&=& u^2\left(
	h(u)\, d\hat t^2+dx_1^2+dx_2^2+dx_3^2\right)
	+\frac{du^2}{h(u) u^2}
	+d\Omega_5^2\ ,
\ee
where $h(u)\equiv 1-{u_0^4}/{u^4}$, $u_0=\pi T$ and $d\Omega_5^2$ is 
the metric on $S^5$.  (Here $\hat t$ is understood to be compact.) 
Defining $z_1$ and $\hat \varphi_2$ so that
\be
dx_1^2+dx_2^2=dz_1^2+z_1^2 d\hat\varphi_2^2\ ,
\ee 
we can perform a TsT transformation on the global $U(1)\times U(1)$
isometry parameterized by $\hat\varphi_2$ and $\hat t$. 
The resulting $\gamma$-deformed metric is 
\be
ds_{T\not=0}^2/R^2&=&u^2\left(
	h(u)G\, dt^2+dz_1^2 G\, d\varphi_2^2 +dx_3^2\right)
	+\frac{du^2}{u^2 h(u)}+d\Omega_5^2\ ,
\ee
where
\be
G=\frac{1}{1+\METgamma^2 u^4 h(u) z_1^2}\ .
\ee

This metric, though obviously different from (\ref{Euc_def_metric}), is somewhat 
similar to the zero-temperature case. 
One salient feature is that the deformation vanishes at the horizon, since $G=1$ when 
$h(u=u_0)=0$.   This fact ensures that thermodynamic properties are not spoiled. 
In particular, the eight-dimensional area of 
the horizon is unchanged, and the surface gravity
\be
\kappa=-\frac{1}{2}(\nabla^a \chi^b)(\nabla_a \chi_b)\ ,
\ee
where $\chi=\frac{\partial}{\partial t}$, 
is also unaffected by the deformation. 
By the first law, the energy is also 
unchanged, though this could be checked directly by calculating the mass in this deformed
background. 
The total number of degrees of freedom is therefore preserved under the deformation. 
We expect that the same holds for the other 
deformations of interest: in the case of the TsT-deformed $S^5$ background, 
this statement corresponds to the total volume being unchanged by the 
deformation.


\section{Classical integrability }
To demonstrate that string theory on our $\gamma$-deformed $AdS_5$ background remains
integrable at the classical level, we formulate the theory in terms of a Lax representation.
As argued by Frolov in \cite{hep-th/0503201}, the crucial issue lies in finding a local Lax 
pair invariant under the $U(1)$ isometry transformations that are generated
as part of the TsT deformation.  
Such a representation was found for the $\gamma$-deformed $S^5$ system in \cite{hep-th/0503201},
and, in describing the analogous computation on the deformed $AdS_5$ geometry,
we will closely follow the treatment therein.  

A useful parameterization of 
the bosonic coset space $(SO(4,2)\times SO(6)) / (SO(5,1)\times SO(5))$ was given 
in \cite{hep-th/0411089}, where the the $AdS_5$ sector takes the form
\be
g = \left(
\begin{array}{cccc}
0		& Z_1		& -Z_3		& \bar Z_2	\\
-Z_1		& 0		& Z_2		& \bar Z_3	\\
Z_3		& -Z_2		& 0		& -\bar Z_1	\\
-\bar Z_2	& -\bar Z_3	& \bar Z_1	& 0				
\end{array}
\right)\ , \qquad Z_i \equiv \eta_i e^{i \hat\varphi_i}\ .
\ee
With the condition in eqn.~(\ref{so21inv}), this matrix is constructed to satisfy
\be
g^\dag\, s\, g = s\ , \qquad s \equiv {\rm diag}(-1,-1,1,1)\ .
\ee
In other words, $g$ is an $SU(2,2)$ embedding of an element of the coset 
$SO(4,2)/SO(5,1)$.  The action for the sigma model is then 
given by that of the principal chiral model
\be
S = \int d\tau d\sigma \gamma^{\alpha\beta} 
	{\rm Tr} \left( g^{-1} \del_\alpha g\, g^{-1} \del_\beta g\right)\ .
\ee

The Lax formulation of an integrable system encodes the (typically nonlinear)
equations of motion in an auxiliary linear problem, which, in turn, is defined
as a flatness condition on conserved currents.
The Lax operator $D_\alpha$ for the sigma model can be written (as a function of
a spectral parameter $x$) as
\be
\label{laxop1}
D_\alpha = \del_\alpha - \frac{j_\alpha^+}{2(x-1)} + \frac{j_\alpha^-}{2(x+1)} 
	\equiv \del_\alpha - {\cal A}_\alpha(x)\ ,
\ee
where $j_\alpha^+$ and $j_\alpha^-$ are self-dual and anti-self-dual projections 
of the right current 
\be
j_\alpha = g^{-1} \del_\alpha g\ ,
\ee
and ${\cal A}_\alpha(x)$ is the (right) Lax connection.
The equations of motion $(\del_\alpha(\gamma^{\alpha\beta} j_\beta) = 0)$ 
are thus encoded by the condition\footnote{To be sure, the auxiliary linear problem is a system
of equations for which eqn.~(\ref{flatness}) stands as a consistency condition.}
\be
\label{flatness}
\left[ D_\alpha, D_\beta \right] = 0\ .
\ee

As noted, the goal is to find a Lax representation invariant under
the TsT deformation that leads to the $\gamma$-deformed $AdS_5$ geometry computed in
eqn.~(\ref{DEFMET}).
As it stands, the Lax operator in eqn.~(\ref{laxop1}) exhibits an explicit
dependence on the $U(1)$ coordinates $\hat\varphi_i$.  Following \cite{hep-th/0503201},
it is straightforward to demonstrate this dependence by noting the factorization
\be
g(\eta_i,\hat\varphi_i) = M(\hat\varphi_i) \tilde g(\eta_i) M(\hat\varphi_i)\ ,
	\qquad M(\hat\varphi_i) = e^{\Phi(\hat\varphi_i)}\ ,
\ee
where the matrix $\Phi$ is given by 
\be
\Phi(\hat\varphi_i) = \frac{i}{2}\left(
\begin{array}{cccc}
\hat\varphi_1 - \hat\varphi_2 + \hat\varphi_3 & 0 & 0 & 0 \\
0 & \hat\varphi_1 + \hat\varphi_2 - \hat\varphi_3 & 0 & 0 \\
0 & 0 & -\hat\varphi_1 + \hat\varphi_2 + \hat\varphi_3 & 0 \\
0 & 0 & 0 & - \hat\varphi_1 - \hat\varphi_2 - \hat\varphi_3
\end{array}
\right)\ ,
\ee
and the $\eta_i$-dependent matrix $\tilde g(\eta_i)$ takes the form
\be
\tilde g(\eta_i) = \left(
\begin{array}{cccc}
0		& \eta_1	& -\eta_3		&   \eta_2	\\
-\eta_1		& 0		& \eta_2		&   \eta_3	\\
\eta_3		& -\eta_2		& 0		& -  \eta_1	\\
-  \eta_2	& -  \eta_3	&   \eta_1	& 0		
\end{array}
\right)\ .
\ee

At this point it is easy to see that the non-derivative dependence of the Lax 
current $j_\alpha$ on the $U(1)$ coordinates $\hat\varphi_i$ can be gauged away:
\be
\label{GT1}
\tilde\jmath_\alpha(\eta_i,\del\hat\varphi_i) = 
	M j_\alpha(\eta_i,\hat\varphi_i) M^{-1}\ .
\ee
This yields the following explicit form:
\be
\tilde\jmath_\alpha = \tilde g^{-1} \del_\alpha \tilde g
	+ \tilde g^{-1} \del_\alpha \Phi \tilde g
	+ \del_\alpha \Phi\ .
\ee
Under this gauge transformation we obtain a suitable Lax operator, 
local under the $\gamma$ deformation described 
above.\footnote{For further details, the reader is
referred to \cite{hep-th/0503201,hep-th/0503192}.}

\subsection{$\Sl(2)_\gamma$ Lax representation}
One remarkable property of the TsT transformations on the $S^5$ subspace
is that both the gauge and string theories neatly exhibit 
$\gamma$-deformed analogues of the classically closed $\su(2)$
subsector.  This is auspicious, as the myriad techniques associated with 
the closure of this sector that have proved 
useful in the undeformed case come to bear in the deformed theory.
Here we are interested in the analogous truncation from the full sigma model on $AdS_5$ 
to a closed $\Sl(2)$ sector (which amounts to a geometrical reduction
to $AdS_3\times S^1$).  It turns out that
the $\gamma$-deformed $AdS_5$ theory indeed admits the same consistent truncation
to a deformed $\Sl(2)_\gamma$ subsector.


A convenient coordinate parameterization in this sector is given 
by the following $SL(2)$ matrix \cite{hep-th/0410105}:
\be
\label{sl2param1}
g = \left(
\begin{array}{cc}
\cos\hat\varphi_1 \cosh\rho + \cos\hat\varphi_2 \sinh\rho & 
		\sin\hat\varphi_1 \cosh\rho - \sin\hat\varphi_2 \sinh\rho \\
-\sin\hat\varphi_1 \cosh\rho - \sin\hat\varphi_2 \sinh\rho & 
		\cos\hat\varphi_1 \cosh\rho - \cos\hat\varphi_2 \sinh\rho
\end{array}
\right)\ .
\ee
Here again, one is faced with the problem of defining a local Lax current invariant 
under $\gamma$ deformations:  the above parameterization clearly exhibits an
explicit linear dependence on the $U(1)$ coordinates $\hat\varphi_i$.
A suitable gauge transformation, analogous to that in eqn.~(\ref{GT1}), can be found
by rewriting eqn.~(\ref{sl2param1}) as
\be
g = e^{\frac{i}{2}(\hat\varphi_1+\hat\varphi_2)\sigma_2}
	e^{\rho\sigma_3} e^{\frac{i}{2}(\hat\varphi_1-\hat\varphi_2)\sigma_2}\ ,
\ee
where $\sigma_i$ are the usual Pauli matrices.
By assigning $M =  e^{\frac{i}{2}(\hat\varphi_1-\hat\varphi_2)\sigma_2}$, 
we obtain the gauge-transformed right current:
\be
\tilde \jmath_\alpha(\eta_i,\del\hat\varphi_i)  & =& M j_\alpha (\eta_i,\hat\varphi_i) M^{-1} 
\nn\\
\kern-10pt
	& = & \left(
\begin{array}{cc}
\del_\alpha \rho & e^{-\rho} ( \del_\alpha\hat\varphi_1\cosh\rho - 
\del_\alpha\hat\varphi_2\sinh\rho ) \\
- e^{-\rho} ( \del_\alpha\hat\varphi_1\cosh\rho + \del_\alpha\hat\varphi_2\sinh\rho ) & 
-\del_\alpha\rho 
\end{array}
\right)\ .
\nn\\
&&
\ee
Invoking the same gauge transformation on the operator $D_\alpha$ yields 
\be
D_\alpha \to  M D_\alpha M^{-1} \equiv \del_\alpha - {\cal R}_\alpha\ , 
\ee
from which one obtains a gauge-transformed Lax connection:
\be
\label{GTlax}
{\cal R}_\alpha  =  M {\cal A}_\alpha M^{-1} - M \del_\alpha M^{-1}
	= \tilde {\cal A}_\alpha + \frac{i}{2}(\del_\alpha\hat\varphi_1 
	- \del_\alpha\hat\varphi_2)\sigma_2\ .
\ee

The auxiliary linear problem defined by the Lax formulation introduces a
monodromy
\be
\Omega(x) = {\cal P} \exp \int_0^{2\pi} d\sigma\, {\cal R}_1(x)\ ,
\ee
and the usual quasi-momentum $p(x)$ is defined in terms of this monodromy as
\be
{\rm Tr}\, \Omega(x) = 2\cos p(x)\ .
\ee
According to the standard argument, the quasi-momentum is conserved (i.e., 
does not depend on $\tau$) because the trace of the holonomy of a flat 
connection does not depend on the contour of integration.   The
dependence of $p(x)$ on the spectral parameter $x$ then implies an infinite
set of conserved integrals of motion (which may be obtained, for example, by
Taylor expansion in $x$).

\subsection{Thermodynamic Bethe equations}
With a suitable Lax representation in hand, we can encode the spectral problem
in a thermodynamic Bethe equation by studying the pole structure and asymptotics
of the quasi-momentum $p(x)$ on the complex $x$ plane.  Here the Bethe ansatz 
appears as a Riemann-Hilbert problem.  
These techniques were developed for the $\su(2)$ sector of the gauge theory
in \cite{hep-th/0402207}, and later applied to the $\Sl(2)$ sector in \cite{hep-th/0410105}.
We will follow these treatments closely, employing the methods presented in \cite{hep-th/0503192} 
for dealing with the general problem of including $\gamma$ deformations. 
In this respect, it turns out to be advantageous 
to work with both the gauge-transformed Lax connection ${\cal R}_\alpha$ in eqn.~(\ref{GTlax})
and the original connection ${\cal A}_\alpha$ appearing in in 
eqn.~(\ref{laxop1}).

To fix boundary conditions in the Riemann-Hilbert problem, one must first study
the structure of the quasi-momentum $p(x)$ in the deformed theory.
Following the analogous argument in \cite{hep-th/0503192}, 
we note that the Lax connections in both the deformed and undeformed sigma models 
can be simultaneously diagonalized by a gauge transformation depending only on 
$\del_\alpha \hat\varphi_i$ (and hence invariant under the deformation).  
This implies that the poles developed by the 
quasi-momentum at $x = \pm 1$ will be the same in both theories:
\be
p(x) = \pi \frac{J/\sqrt{\lambda} \mp m}{x\pm 1} + \cdots \qquad x \to \mp 1\ ,
\ee
where $m$ is the winding number around the decoupled $S^1$ in $AdS_3 \times S^1$
(classical strings in the $\Sl(2)$ subsector propagate on this subspace).

At $x=0$ and $x=\infty$ the quasi-momentum develops additional constant 
contributions due to the deformation.  In these cases it is helpful to rely on 
the asymptotics of the original Lax connection ${\cal A}_\alpha$ by invoking an
inverse gauge transformation on the monodromy \cite{hep-th/0503192}:
\be
T(x) = M(2\pi) {\cal P} \exp \int_0^{2\pi} d\sigma {\cal A}_1(x) M^{-1}(0)\ .
\ee
This yields the following representation of the quasi-momentum:
\be
\label{GTmono}
2\cos p(x) &=& {\rm Tr}\ M_R\, {\cal P} \exp \int_0^{2\pi} d\sigma {\cal A}_1(x)\ , 
\\
{\cal A}_1(x) & = & \frac{j_1}{x^2-1} + \frac{x\, j_0}{x^2-1}\ ,
\ee
where the matrix $M_R$ is given by
\be
M_R = M^{-1}(0) M(2\pi) = \left(
\begin{array}{cc}
\cos  \Mgamma\pi (S-\Delta) & -\sin \Mgamma\pi (S-\Delta) \\
\sin  \Mgamma\pi (S-\Delta) & \cos \Mgamma\pi (S-\Delta)
\end{array}
\right)\ .
\ee
We have made use here of the twisted boundary conditions in eqn.~(\ref{TBC}) above,
where we identify the $U(1)$ charges $J_1$ and $J_2$ with the energy $\Delta$ and 
impurity number $S$ of corresponding string energy eigenstates according to
\be
J_1 = -\Delta\ , \qquad J_2 = S\ .
\ee
This notation is borrowed from the undeformed $\Sl(2)$ sector, where
$\Delta$ and $S$ are respectively mapped to the dimension and spin of corresponding 
$\Sl(2)$ operators in the gauge theory.

By parameterizing the right and left currents of the sigma model in a standard fashion 
\be
j_\alpha = g^{-1} \del_\alpha g = \frac{1}{2}j_\alpha \cdot \hat\sigma\ ,
\qquad
l_\alpha =  \del_\alpha g g^{-1} = \frac{1}{2}l_\alpha \cdot \hat\sigma\ ,
\ee
where $\hat\sigma = (i\sigma_2,\sigma_3,-\sigma_1)$, one may refer 
directly to the undeformed $\Sl(2)$ problem \cite{hep-th/0410105} to compute
\be
\frac{\sqrt{\lambda}}{4\pi}\int_0^{2\pi}d\sigma\, j_0^0 = \Delta +S\ ,
\qquad
\frac{\sqrt{\lambda}}{4\pi}\int_0^{2\pi}d\sigma\, l_0^0 = \Delta -S\ .
\ee
The asymptotics of the quasi-momentum at $x=0$ and $x=\infty$ can thus be found 
by expanding eqn.~(\ref{GTmono}) in the spectral parameter 
and following the prescription provided in \cite{hep-th/0503192} 
for discarding nonlocal contributions.\footnote{Adopting the methodology in \cite{hep-th/0503192} 
appears to generate sensible answers for the $\Sl(2)_\gamma$ sector,
despite the possible ambiguities noted therein.}
We find 
\be
\label{pBC}
p(x) &=& \pi\Mgamma (\Delta-S) + 2\pi \frac{\Delta+S}{\sqrt{\lambda}\, x} + \cdots\ , \qquad 
	x \to \infty\ ,
\nn\\
p(x) &=& \pi\Mgamma (\Delta+ S) - 2\pi \frac{\Delta-S}{\sqrt{\lambda}} x + \cdots\ , \qquad 
	x \to 0\ .
\ee
A resolvent function analytic on the complex $x$ plane can therefore be defined by subtracting from the 
quasi-momentum its poles at $x=\pm 1$ and the constant contribution $ \pi\Mgamma (\Delta-S)$
at $x \to \infty$:
\be
G(x) = p(x) - \pi\frac{J/\sqrt{\lambda}+m}{x-1} - \pi\frac{J/\sqrt{\lambda}-m}{x+1} 
	- \pi\Mgamma (\Delta - S)\ .
\ee
(Subtracting the constant piece serves to allow sensible Cauchy integrals over the spectral
density, defined below.)
From eqn.~(\ref{pBC}), one obtains the following asymptotic behavior 
\be
G(x) & = & \frac{2\pi}{\sqrt{\lambda}\, x}(\Delta+ S - J) + \cdots\ , \qquad x\to \infty \nn\\
G(x) & = & 2\pi (m + \Mgamma  S) + \frac{2\pi x}{\sqrt{\lambda}}(S-\Delta+J) + \cdots\ ,
	\qquad x\to 0\ .
\ee

The usual spectral representation of the resolvent takes the form
\be
\label{spectralG}
G(x) = \int_C d{x'} \frac{\sigma({x'})}{x-{x'} }\ , \qquad C = C_1 \cup C_2 \ldots \cup C_n\ ,
\ee
where $\sigma(x)$ stands as a spectral density function supported on a finite number of cuts $C_i$ in 
the complex $x$ plane.  Given that (by construction) $G(x)$ is analytic in the spectral parameter,
it is straightforward to derive the following constraint equations on $\sigma(x)$:
\be
\label{RHBC}
\int_C d{x}\, \sigma({x}) & = & \frac{2\pi}{\sqrt{\lambda}}(\Delta+S-J)\ , \nn\\
\int_C d{x}\, \frac{\sigma({x})}{{x}} & = & - 2\pi (m + \Mgamma  S)\ , \nn\\
\int_C d{x}\, \frac{\sigma({x})}{{x}^2} & = & \frac{2\pi}{\sqrt{\lambda}}(\Delta-S-J)\ .
\ee
Unimodularity of the monodromy $\Omega(x)$ implies 
\be
p(x+i0) + p(x-i0) = 2\pi n_k\ , \qquad x \in C_k\ ,
\ee
where the mode integer $n_k$ labels the set of eigenvalues supported in the $k^{\rm th}$
contour $C_k$:  this number corresponds to the mode number of individual impurity
excitations on the string worldsheet.
One thereby obtains the finite-gap integral equation
\be
\label{RH1}
2\pint_{\kern-5pt C}
	 d{x'}\, \frac{\sigma({x'})}{x-{x'}} = -2\pi\left(
	\frac{J/\sqrt{\lambda}+m}{x-1} + \frac{J/\sqrt{\lambda}-m}{x+1}
	\right)
	+ 2\pi n_k
	- 2\pi\Mgamma (\Delta - S)\ ,
\ee
where, as usual, $x$ is understood in this context 
to take values in $C_k$.  This equation represents a thermodynamic Bethe ansatz in 
the classical limit of the deformed string theory in the $\Sl(2)_\gamma$ sector.
As expected, the $\Mgamma\to 0$ limit of this 
equation reduces to the original Riemann-Hilbert problem obtained for the undeformed 
$\Sl(2)$ sector in \cite{hep-th/0410105}.  

Compared to the corresponding result in the deformed $\su(2)_\gamma$ sector, we have obtained
a relatively complicated modification to the undeformed $\Sl(2)$ thermodynamic Bethe equation.
Using the constraints in eqns.~(\ref{RHBC}), we can rewrite eqn.~(\ref{RH1}) 
in a slightly more illuminating form:
\be
\label{RH2}
&&
	2\pi(n_k - \Mgamma J)
	-4\pi\frac{x\, J/\sqrt{\lambda} }{x^2-1} 
	\,\, = \,\,
\nn\\
&&\kern+50pt
	2\pint_{\kern-5pt C} dx'\,\sigma(x')\left(
	\frac{1}{x-x'} 
	-\frac{2x' + \Mgamma\sqrt{\lambda}\, ({x'}^2-1)}{2\, {x'}^2(x^2-1)}
	+ \frac{\Mgamma\sqrt{\lambda}}{2}\, \frac{1}{{x'}^2}
	\right)\ .
\ee
In addition to a global shift in the mode number $n_k$,
several $\Mgamma$-dependent deformation terms now appear under the spectral 
integral. 
An obvious question at this point is whether the established 
technology for promoting thermodynamic Bethe equations to {\it discrete}, 
(or ``quantum'') Bethe equations will be completely reliable for string theory in this 
wider class of $\gamma$-deformed backgrounds.
In the $\su(2)_\gamma$ sector, the TsT deformation simply amounts to a 
shift by $\Mgamma J$ in the mode number $n_k$:
the $\su(2)_\gamma$ problem is therefore solved trivially by invoking a corresponding shift 
in the known $\su(2)$ quantum string Bethe ansatz.
While we have presently obtained the same shift 
in the $\Sl(2)_\gamma$ sector of the string theory, we have also generated
additional $\Mgamma$-dependent contributions on the right-hand side of eqn.~(\ref{RH2}), 
which mark an interesting nontrivial deformation of the problem.  
We will return to the crucial issue of deformed quantum string Bethe 
ans\"atze in section \ref{quantumbethe}.

\subsection{General deformations}
To some extent we have circumvented the problem of T-duality along timelike
directions.  We are forced to confront this issue, however, if we wish to consider
a wider class of deformations achieved by sequential TsT transformations on 
all of the $U(1)$ angular coordinates $\hat\varphi_i$ in $AdS_5$.  It turns out that, with respect to
the goals set forth in the present study, it is efficient to adopt a pragmatic viewpoint by
invoking deformations in a strictly formal manner.  If one prefers, 
the resulting background may then be studied at face-value, and not necessarily as a deformation 
of any particular parent geometry.

Following \cite{hep-th/0503201}, we obtain a three-parameter deformation of $AdS_5$
using a chain of TsT transformations on each of the three tori in $AdS_5$ parameterized by the 
undeformed $U(1)$ angular coordinates
$(\hat\varphi_1,\hat\varphi_2)$, $(\hat\varphi_2,\hat\varphi_3)$ and $(\hat\varphi_1,\hat\varphi_3)$:
\be
ds^2/R^2 &=& g^{ij} d\eta_i d\eta_j  
		+ G ( -\eta_1^2 d\varphi_1^2 
		+ \eta_2^2 d\varphi_2^2 
		+ \eta_3^2 d\varphi_3^2 ) 
	- G  \eta_1^2\eta_2^2\eta_3^2 \bigl[ d (\sum^3_{i=1} \METgamma_i \varphi_i)\bigr]^2\ ,
\nn\\
B_2 &=&  R^2  G w_2 \ , 
\nn\\
w_2  &\equiv&  -\METgamma_3\,  \eta_1^2 \eta_2^2\, d\varphi_1 \wedge d\varphi_2 +
\METgamma_1\,  \eta_2^2 \eta_3^2\, d\varphi_2 \wedge d\varphi_3 - \METgamma_2\,
\eta_3^2 \eta_1^2\, d\varphi_3 \wedge d\varphi_1\ ,  
\ee
with 
\be
G^{-1} &\equiv& 1 - \METgamma_3^2\,  \eta_1^2 \eta_2^2
	+ \METgamma_1^2\,  \eta_2^2 \eta_3^2
	- \METgamma_2^2\,  \eta_1^2 \eta_3^2\ .
\ee

Starting from this geometry, one may proceed according to the methodology described
above.   As with the $\su(2)_\gamma$
sectors described in section \ref{su2DEF}, we find that each individual $\Sl(2)_\gamma$
sector, to $O(1/R^2)$ in the large-radius expansion, is deformed by a single 
element of the set $\{\METgamma_i \}$.  
Following section \ref{AdSDEF}, one obtains the following
transformation conditions under the full sequence of TsT deformations, analogous to
eqn.~(\ref{TBC0}) above:
\be
\hat\varphi_1' & \to  &  \varphi_1' - \Mgamma_3\, p_2 - \Mgamma_2\, p_3\ , \nn\\
\hat\varphi_2' & \to  &  \varphi_2' + \Mgamma_1\, p_3 + \Mgamma_3\, p_1\ , \nn\\
\hat\varphi_3' & \to  &  \varphi_3' + \Mgamma_2\, p_1 - \Mgamma_1\, p_2\ .
\ee
(We again introduce the modified deformation parameters $\Mgamma_i = \METgamma_i/\sqrt{\lambda}$.) 
We note that the sign of $\Mgamma_3$ is a consequence of the 
$\hat\varphi_2,\ \hat\varphi_1,\ \hat\varphi_2$ 
ordering of the corresponding TsT transformation; this sign would be reversed if 
we instead chose the sequence $\hat\varphi_1,\ \hat\varphi_2,\ \hat\varphi_1$
(the other transformations are ordered as in \cite{hep-th/0503201}: 
$\hat\varphi_2,\ \hat\varphi_3,\ \hat\varphi_2$ and $\hat\varphi_1,\ \hat\varphi_3,\ \hat\varphi_1$).
We therefore obtain the following
twisted boundary conditions on the undeformed $U(1)$ coordinates $\hat\varphi_i$ that 
arise under this chain of TsT deformations (again, $m_i$ stand for winding numbers):
\be
\hat\varphi_1(2\pi) - \hat\varphi_1(0) & = & 2\pi(m_1 - \Mgamma_3\, J_2 - \Mgamma_2\, J_3)\ , \nn\\
\hat\varphi_2(2\pi) - \hat\varphi_2(0) & = & 2\pi(m_2 + \Mgamma_1\, J_3 + \Mgamma_3\, J_1)\ , \nn\\
\hat\varphi_3(2\pi) - \hat\varphi_3(0) & = & 2\pi(m_3 + \Mgamma_2\, J_1 - \Mgamma_1\, J_2)\ .
\ee
With these boundary conditions, it is easy to follow the above procedures to obtain
a Lax representation and thermodynamic Bethe ansatz in this more general deformed geometry.
Since truncation to $\Sl(2)_\gamma$ subsectors restricts to a single $\Mgamma_i$ deformation,
however, questions pertaining to these sectors can be studied by choosing a single
TsT transformation.  (In the coordinate system employed in this paper, the deformation 
parameterized by $\Mgamma_1$ is in fact trivial upon truncation to an $\Sl(2)_\gamma$ subsystem.)

\section{Twisted string spectra in the near-pp-wave limit}
We now turn to the task of gathering data on the spectrum of string states in 
protected sectors of the string theory on $\gamma$-deformed $AdS_5\times S^5$.  
For arbitrary numbers of worldsheet excitations, with arbitrary subsets of bound states
(marked by subsets of confluent mode numbers), formulas for the perturbative
$O(1/J)$ energy correction away from the pp-wave limit are typically complicated, and provide
a fairly rigorous test of any conjectural Bethe equations that purport to encode such
spectral information \cite{hep-th/0407240}.
In this section we will compute these near-pp-wave corrections in the deformed $\su(2)_\gamma$ and 
$\Sl(2)_\gamma$ sectors described above.

\subsection{$\su(2)_\gamma$ sector}
Since we are working within bosonic truncations of the full superstring theory 
on the $\gamma$-deformed geometry, we find it convenient to approach the 
computation of energy spectra using a purely Hamiltonian 
formalism.\footnote{See, e.g., \cite{hep-th/0505028} for a detailed description of 
such an approach in the undeformed $AdS_5\times S^5$ background.}
One particular advantage of such a framework is that components of the 
worldsheet metric may be employed as Lagrange multipliers enforcing the Virasoro
constraints.  As such, we need not compute curvature corrections to the worldsheet
metric that are inevitable in other approaches.  Omitting the computational details,
we find the following lightcone Hamiltonian, truncated to an $\su(2)_\gamma$ sector on the
deformed $S^5$ subspace:
\be
H_{\rm LC} = \frac{1}{G^{++}p_-}\left(
	G^{+-}\tilde p_- p_- - \sqrt{F} \right)
	-\frac{p_A {x'}^A}{p_-}B_{+-}
	+B_{+y} y'
	+B_{+\bar y}\bar y'\ .
\ee
For the sake of compactness, we have defined the quantity
\be
F & \equiv & (G^{+-})^2 \tilde p_-^2 p_-^2 - G^{++} \Bigl[
	G_{--}(p_A {x'}^A)^2
	+p_-^2 \Bigl(
	\tilde p_y G^{yy} \tilde p_y
	+ 2 \tilde p_y G^{y \bar y} \tilde p_{\bar y} 
	+ \tilde p_{\bar y} G^{\bar y \bar y} \tilde p_{\bar y}
\nn\\
&&	+ G^{--}\, \tilde p_-^2
	+ y' G_{y y} y' 
	+ 2 y' G_{y \bar y} \bar y'
	+ \bar y' G_{\bar y \bar y} \bar y'
	\Bigr) \Bigr]\ .
\ee
The notation $\tilde p_-$ and 
$\tilde p_y$ is used to indicate corrections to the
usual conjugate momenta $p_-$ and $p_y$ due to the presence of a nonzero $B$-field:
\be
\tilde p_- &\equiv& p_- + B_{-y} y' + B_{-\bar y} \bar y'\ ,
\nn\\
\tilde p_y &\equiv& p_y + B_{y\bar y}\bar y' 
	+ B_{-y}\frac{p_A {x'}^A}{p_-}\ .
\ee

The vector $x^A$ is understood to span the coordinate set 
$(x^+,x^-,y,\bar y)$, where the complex fields $y$ and $\bar y$ were defined in 
eqn.~(\ref{complex}) above.  The restriction to the complex pair $(y,\bar y)$ (as opposed
to the $(z,\bar z)$ pair from eqn.~(\ref{complex})) corresponds to the truncation 
from the full theory on the $\gamma$-deformed $S^5$ to an $\su(2)_\gamma$ 
sector.\footnote{Technically, the theory restricted to $(y,\bar y)$ coordinates 
describes a system that is slightly larger than the closed $\su(2)_\gamma$ sector.
To fully truncate to $\su(2)_\gamma$ we will perform an additional projection
that is described below.}
In the corresponding $\Sl(2)_\gamma$ truncation on the deformed $AdS_5$ subspace
we will project onto coordinates $(x^+,x^-,v,\bar v)$, where $v$ and $\bar v$ are defined in 
eqn.~(\ref{complex2}).

We aim to compute near-pp-wave energy spectra in a semiclassical expansion about point-like 
(or BMN) string solutions.
(Note that analogous corrections were found for a different set of string solutions 
in an $\su(2)_\gamma$ sector in \cite{hep-th/0503192}.)
Arranging the large-radius (equivalently, large-$J$) expansion of the 
lightcone $\su(2)_\gamma$ Hamiltonian according to
\be
H_{\rm LC} = H_0 + \frac{H_{\rm int}}{ R^2 } + O(1/R^4)\ ,
\ee
we obtain the following as functions of coordinate fields 
on the $\gamma$-deformed $S^5$ subspace:
\be
H_0(S^5_\METgamma)
	 &=& \frac{1}{2p_-}\Bigl[
	4 |p_y|^2 + |y'|^2 - i p_-(y' \bar y - y \bar y')\METgamma
	+ p_-^2 |y|^2 (1+\METgamma^2) \Bigr]\ ,
\nn\\
{H}_{\rm int}(S^5_\METgamma)
	 &=& \frac{1}{8  p_-^3 }\Bigl\{
	-4p_y^2 (4 \bar p_y^2 + p_-^2 y^2 - {y'}^2)
	-16p_-^2|p_y|^2|y|^2
	+p_-^2 \bar y^2 (3p_-^2 y^2 + {y'}^2 - 4\bar p_y^2)
\nn\\
&&
	+ \bar {y'}^2 (p_-^2 y^2 - {y'}^2 + 4\bar p_y^2)
	-2 i p_-\METgamma\Bigl[
	-4 p_y^2 y y' + p_-^2 |y|^2(y \bar y' - \bar y y')
\nn\\
&&
	+\bar y' (4 \bar p_y^2 \bar y - {y'}^2 \bar y + y |y'|^2)
	\Bigr]
	- p_-^2 \METgamma^2 \Bigl[
	4 p_y^2 y^2 + \bar y^2(4 \bar p_y^2 + 2 p_-^2 y^2 - {y'}^2)
	+4 |y|^2 |{y'}|^2 
\nn\\
&&
	- y^2 \bar {y'}^2 
	+ 2 i p_- \METgamma |y|^2 (y\bar y' - y'\bar y)
	+ p_-^2\METgamma^2 |y|^4 
	\Bigr]
	\Bigr\}\ .
\ee
As expected, the pp-wave Hamiltonian is quadratic in worldsheet fluctuations, while
the interaction Hamiltonian $H_{\rm int}$ appearing at $O(1/R^2)$ in the expansion 
contains terms that are uniformly quartic in fields.

The leading-order equations of motion are solved by the usual expansion in 
Fourier modes\footnote{Note that, strictly speaking,
one is instructed in this formalism to compute equations of motion directly from the 
Hamiltonian rather than the Lagrangian.  The difference between the two formalisms 
amounts to a sign flip on the deformation parameter $\METgamma$.}
\be
y(\tau,\sigma) = \sum_{n=-\infty}^\infty y_n(\tau) e^{-i n \sigma }\ ,
	\qquad 
p(\tau,\sigma) = \sum_{n=-\infty}^\infty p_n(\tau) e^{-i n \sigma }\ ,
\ee
where
\be
y_n(\tau) = \frac{i}{\sqrt{2\omega_n}} 
	(a_n e^{-i\, \omega_n \tau} - \bar a_{-n}^\dag e^{i\, \omega_n \tau} )\ ,
	\qquad
p_n(\tau) = \frac{1}{2}\sqrt{ \frac{\bar\omega_n}{2}} 
	(\bar a_n e^{-i\, \bar\omega_n \tau} +  a_{-n}^\dag e^{i\, \bar\omega_n \tau} )\ ,
\ee
and $n$ denotes an integer mode index $(-\infty < n < \infty)$.
In the presence of a nonzero deformation, we obtain the following shifted 
dispersion relations:
\be
\label{omegadef}
\omega_n^2 = p_-^2 + (n-p_-\METgamma)^2\ , \qquad 
\bar\omega_n^2 = p_-^2 + (n+p_-\METgamma)^2\ .
\ee
Upon expanding the interaction Hamiltonian in raising and lowering operators,
we complete the projection onto the closed $\su(2)_\gamma$ 
sector by keeping either the $(a_n,a_{-n}^\dag)$ or $(\bar a_n,\bar a_{-n}^\dag)$ 
oscillator pair and setting all remaining terms to zero.
In essence, this achieves an $SO(4)$ symmetric-traceless projection, in
precise analogy with the undeformed theory 
\cite{hep-th/0307032,hep-th/0404007,hep-th/0407240,hep-th/0405153}.
The Hamiltonian is then understood to block-diagonalize in the Fock subspace 
spanned by $N$-impurity string states
composed of $N$ raising operators in the $\su(2)_\gamma$ projection acting 
on a ground state labeled by $\ket{J}$:
\be
a_{n_1}^\dag a_{n_1}^\dag \cdots a_{n_N}^\dag \ket{J}\ .
\nn
\ee

Following the usual conventions, it is convenient to 
replace the $S^5$ radius $R$ with $\sqrt{p_- J}$ and arrange the large-$J$ expansion of the
energy spectrum according to
\be
E(\{n_j\},J) = \sum_{j=1}^N \sqrt{1+(n_j\ -\METgamma/\sqrt{\lambda'})^2 \lambda'}
	+ \delta E(\{n_j\},J) + O(1/J^2)\ .
\ee
The leading-order term in this expansion 
represents the familiar BMN energy formula with an
additional $\METgamma$-dependent shift in the mode index.
The $N$ worldsheet excitations can be labeled
by $N$ integer mode numbers $n_j$  
such that the full set $\{n_j\}$ is populated
by $M$ uniform subsets consisting of $N_{j}$ equal mode 
numbers $n_j$ $(j \in 1,\ldots ,M)$:\footnote{Loosely speaking, one can think of each 
subset of $N_j$ equal mode numbers 
as corresponding to a bound state on the string worldsheet.  These states
in turn fill out the support contours $C_i$ in the complex plane of the spectral 
parameter $x$.}  
\be
\{n_j\} = 
	\Bigl\{
	\{ \underset{N_{1}}{\underbrace{ n_1,n_1,\ldots ,n_1 }} \},
	\{ \underset{N_{2}}{\underbrace{ n_2,n_2,\ldots ,n_2 }} \},
	\ldots ,
	\{ \underset{N_{M}}{\underbrace{ n_M,n_M,\ldots ,n_M }} \}
	\Bigr\}\ .
\ee
The perturbative energy shift in the near-pp-wave limit is then given by
\be
\label{su2shift}
\delta E_{\su(2)_\gamma} (\{n_j\},\{N_{j}\},J) & = &
	-\frac{1}{2J}\biggl\{
	\sum_{j=1}^M N_{j} (N_{j}-1) 
	\left[\left(1+(\METgamma - n_j\sqrt{\lambda'})^{-2}\right)^{-1} \right]
\nn\\
&&
\kern-40pt
	- \sum_{j,k = 1 \atop j \neq k}^M \frac{N_{j} N_{k}}{\omega_{n_j}\omega_{n_k}\lambda'}
	\Bigl\{  -\lambda'(n_j n_k + n_k^2 + n_j^2(1+n_k^2 \lambda'))
\nn\\
&&	
\kern-40pt
	+ \METgamma ((n_j+n_k)\sqrt{\lambda'}-\METgamma)
	(3 + 2n_j n_k \lambda' - (n_j + n_k)\sqrt{\lambda'}\METgamma + \METgamma^2)
\nn\\
&&
\kern-40pt
	+ \lambda' (n_j\sqrt{\lambda'}-\METgamma)(n_k\sqrt{\lambda'}-\METgamma)\omega_{n_j}\omega_{n_k} 
	\Bigr\} \biggr\}\ .
\ee

At this point it is useful to recall (see eqn.~(\ref{gammascale})) that the fixed 
parameter $\METgamma$ appearing in the geometry is related to the
corresponding deformation parameter in the $\beta$-deformed gauge theory by
$\beta = \Mgamma = \METgamma/\sqrt{\lambda} $. As noted in \cite{hep-th/0503192},
this implies that the parameter
\be
\tilde \beta \equiv \beta J \sqrt{\lambda'}
\ee
is also held fixed in the large-$J$ expansion about the pp-wave limit.
Since $\lambda'$ is fixed and finite in this limit, $\beta J$ must also
be held fixed.  
The above formula for the energy shift at $O(1/J)$ thus has a very simple 
interpretation: if we take the undeformed near-pp-wave energy correction \cite{hep-th/0407240}
\be
\delta E_{\su(2)}(\{n_j\},\{N_{j}\},J) & = & -\frac{1}{2J}\biggl\{
	\sum_{j=1}^M N_{j}(N_{j}-1)
	\left(1-\frac{1}{\varpi_{n_j}^2\lambda'}\right)
\nn\\
&&\kern-18pt
	+\sum_{j,k=1\atop j\neq k}^M \frac{N_{j}N_{k}}{\varpi_{n_j}\varpi_{n_k}}
	\left[q_k^2 + q_j^2\varpi_{n_k}^2\lambda'
	+q_j q_k(1-\varpi_{n_j}\varpi_{n_k}\lambda')\right]\biggr\}
\label{SU2FULL}
\ee
(the symbol $\varpi_n$ is specified by the undeformed dispersion relation
$\varpi_n=\sqrt{p_-^2 + n^2}$),
and shift the mode numbers by the fixed amount 
$n_j\rightarrow n_j - \beta J$, we obtain eqn.~(\ref{su2shift})
exactly.  Based on observations made in \cite{hep-th/0503201,hep-th/0503192},
this is precisely the outcome one should expect:  the deformed theory is mapped from the
original theory on $S^5$ by imposing twisted boundary conditions, analogous to
those in eqn.~(\ref{TBC}), on the relevant undeformed fields.  In the $\su(2)_\gamma$
sector, these boundary conditions only act to shift the mode numbers of the worldsheet
excitations.   We shall see a more dramatic modification in the $\Sl(2)_\gamma$ sector.

\subsection{$\Sl(2)_\gamma$ sector}
The near-pp-wave limit taken in the $\Sl(2)_\gamma$ sector yields the
following string lightcone Hamiltonian, expanded to $O(1/J)$ near the pp-wave limit 
and projected onto the complex coordinate
pair $(v,\bar v)$:
\be
H_0(AdS_5^\METgamma)
	 &=& \frac{1}{2p_-}\Bigl[
	4 |p_v|^2 + |v'|^2 - i p_-(v' \bar v - v \bar v')\METgamma
	+ p_-^2 |v|^2 (1+\METgamma^2) \Bigr]\ ,
\nn\\
H_{\rm int}(AdS_5^\METgamma)
	 & = & \frac{1}{8p_-^3}\biggl\{
	16 p_-^2 |p_v|^2 |v|^2
	+ (4 \bar p_v^2 - v'^2)\bar v'^2
	+ 4 i p_-^3 |v|^2(v \bar v' - v' \bar v)\METgamma (1+\METgamma^2)
\nn\\
&&
\kern-40pt
	+ p_-^4 |v|^4(-1+6\METgamma^2+3\METgamma^4)
	+ 4 p_v^2(-4 \bar p_v^2 + v'^2 + p_-^2 v^2 (1+\METgamma^2) )
\nn\\
&&\kern-40pt
	+p_-^2\Bigl[
	4 |v|^2 |v'|^2 \METgamma^2 
	+ 4 \bar p_v^2 \vc^2 (1+\METgamma^2)
	-v'^2 \vc^2(1+\METgamma^2)
	-v^2 \vc'^2 (1+\METgamma^2) \Bigr]
	\biggr\}\ .
\ee
Expanding in raising and lowering operators, we again project onto $\Sl(2)_\gamma$ sectors
by setting either the $(a_n,a_{-n}^\dag)$ or $(\bar a_n,\bar a_{-n}^\dag)$
oscillator pair to zero.  The perturbing Hamiltonian $H_{\rm int}(AdS_5^\METgamma)$
is then easily diagonalized in a corresponding basis of Fock states to yield
the following energy shift at $O(1/J)$ in the near-pp-wave limit: 
\be
\label{nearppSL2}
\delta E_{\Sl(2)_\gamma} (\{n_j\},\{N_{j}\},J) & = &
	\frac{1}{2J}\biggl\{
	\sum_{j=1}^M N_{j} (N_{j}-1) 
	\frac{ (\METgamma - n_j\sqrt{\lambda'})^2}{\omega_{n_j}^2\lambda'}
\nn\\
&&
\kern-80pt
	+ \sum_{j,k = 1 \atop j \neq k}^M \frac{N_{j} N_{k}}{\omega_{n_j}\omega_{n_k}\lambda'}
	\Bigl\{ 3\METgamma^2 + \METgamma^4 - (n_j+n_k)\METgamma^3\sqrt{\lambda'}
	+ n_j n_k\lambda' (1-n_j n_k\lambda')
\nn\\
&&
\kern-40pt
	+ (n_j +n_k)\METgamma\sqrt{\lambda'}(n_j n_k \lambda'-2)
	+ \lambda'(n_k n_j\lambda' - \METgamma^2)\omega_{n_j}\omega_{n_k}
	\Bigr\}
	\biggr\}\ .
\ee
In this sector the $O(1/J)$ energy spectrum is 
{\it not} related to the corresponding undeformed spectrum by a simple shift in the 
mode numbers.  A closer inspection of eqn.~(\ref{nearppSL2}) reveals, however, that
the deformed energy shift can be understood to arise from the combined effect
of an overall shift in worldsheet mode numbers and an additional shift linear in 
the deformation parameter $\Mgamma$.  Schematically, one finds that
\be
\delta E_{\Sl(2)_\gamma}(\{n_j\}) & = & \delta E_{\Sl(2)}(\{n_j + \Mgamma J\})
		+ \Mgamma\, \delta E_2(\{n_j + \Mgamma J\})\ ,
\ee
where $\delta E_2$ is similar, but not identical, to the undeformed near-pp-wave
energy shift $\delta E_{\Sl(2)}$ from \cite{hep-th/0407240}.
In the next section we will study how these interesting modifications can be 
embedded in a discrete extension of the thermodynamic $\Sl(2)_\gamma$ Bethe ansatz 
computed in eqns.~(\ref{RH1},\ref{RH2}).

\section{Twisted quantum Bethe equations}
\label{quantumbethe}
As described above, the integral equations for the spectral density of the Lax operator 
in eqns.~(\ref{RH1}) and (\ref{RH2}) can be interpreted as Bethe equations for the string
theory in a classical, thermodynamic limit.  
This picture has led to a series of conjectures
for how to discretize this limit of the ``stringy'' Bethe equations based on 
similarities with corresponding limits of the dual gauge theory 
\cite{hep-th/0406256,hep-th/0412188,hep-th/0504190}.  
One remarkable outcome is that the resulting discrete Bethe equations exactly reproduce
near-pp-wave string energy spectra in various closed sectors of the theory.
It is not entirely clear
whether these techniques are specific to the duality connecting ${\cal N}=4$ SYM theory
with string theory on $AdS_5\times S^5$.  To pose the question more precisely, we ask
if the discretization conjectures based on established gauge theory considerations 
can be applied with similar success in other contexts.  Since the gauge theory side
of the duality seems to be drastically modified under deformations corresponding 
(holographically) to 
TsT transformations on the $AdS_5$ subspace, the $\gamma$-deformed $\Sl(2)_\gamma$ 
sectors described here present an excellent opportunity to address such issues.  
In this section we will therefore apply familiar discretization
techniques to the $\gamma$-deformed string theory on $AdS_5$ to derive a twisted
quantum string Bethe ansatz for the spectral problem in these sectors.
While we will briefly review the essential methodology, the reader is referred to 
\cite{hep-th/0406256,hep-th/0412188,hep-th/0504190}
for further details.

The central conjecture is that the string theory 
spectrum is described by the diffractionless scattering of elementary excitations 
on the worldsheet \cite{hep-th/0412188}.  In other words, the energy spectrum should be 
encoded in a fundamental equation in the excitation momenta $p_k$ (and corresponding
mode numbers $n_k$) of the form
\be
\label{theta1}
p_k J = 2\pi n_k + \sum_{ j\neq k} \theta(p_k,p_j)\ ,
\ee
where the scattering phase $\theta(p_k,p_j)$ is defined in terms of a factorized 
$S$-matrix:\footnote{For present purposes we only display the bosonic version of this equation.}
\be
\theta(p_k,p_j) = -i \log S(p_k,p_j)\ .
\ee
This two-body scattering matrix has been the locus of a great deal of deserved attention:
the symmetry algebra of the theory constrains the form of the $S$ matrix
up to an overall phase \cite{hep-th/0511082}, and it is suspected that this phase is further 
constrained by unitarity and crossing symmetry \cite{hep-th/0603038} (or certain worldsheet versions
thereof).  (For additional interesting
developments, see \cite{hep-th/0604135,hep-th/0604175}.)  One certainly hopes that further insight
into this problem will reveal a precise procedure by which the structure of the
string theory $S$ matrix might be uniquely determined.  
The second ingredient, which is key in the present scenario, 
is that the hidden local charges in the theory, labeled
as $Q_r$, are expected to arise as linear sums over local dispersion relations $q_r(p_k)$:
\be
\label{disp}
Q_r = \sum_k q_r(p_k)\ .
\ee

Starting from the classical Bethe equations provided by the Lax representation of the 
string sigma model, discretized equations may be formulated by relying on cues provided
by the gauge theory \cite{hep-th/0406256,hep-th/0412188}.  
One crucial test of this procedure 
is whether predictions obtained from the conjectured quantum Bethe equations 
match data coming directly from string theory computations.  For example,
the near-pp-wave energy shifts in the undeformed $\Sl(2)$ sector 
should be encoded in the (discretized) scattering phase 
$\theta(p_k,p_j)$ \cite{hep-th/0412188}\footnote{This particular equation 
holds for the $\Sl(2)$ sector only.}
\be
\delta\Delta(n_k,n_j,\Mgamma) 
	= \lambda' \sum_{j,k=1\atop j\neq k}^S 
	 \frac{J}{2\pi}\frac{n_k}{\sqrt{1+\lambda' n_k^2}}\,
	\theta\left({2\pi}n_k/J,{2\pi}n_j/J \right)\ .
\ee
Amazingly, this general approach yields the correct energy spectrum in the undeformed theory
at $O(1/J)$ for the closed $\su(2)$, $\su(1|1)$ and $\Sl(2)$ sectors, and for a variety of different
string solutions (for details and further interesting developments, see
\cite{hep-th/0406256,hep-th/0412188,hep-th/0504190,hep-th/0412072,hep-th/0507189,hep-th/0509084,hep-th/0509096,hep-th/0603204,hep-th/0604069}).

As noted above, classical string Bethe equations were derived in the TsT-deformed 
$\su(2)_\gamma$ sector in \cite{hep-th/0503192,hep-th/0503201}.  As expected, these
equations differed from those in the undeformed $\su(2)$ sector by a simple global shift 
in the mode numbers $n_j,\ n_k$.  The quantum extension of the $\su(2)_\gamma$
Bethe equations is therefore rather simple, and the result manifestly agrees with 
the corresponding $\su(2)_\gamma$ energy spectrum at $O(1/J)$, computed in eqn.~(\ref{su2shift}) 
above. (Recall that this energy shift may indeed be obtained from the undeformed spectrum
by an overall shift in mode numbers.)  An analogous treatment
of the thermodynamic Bethe equations in the $\Sl(2)_\gamma$ sector requires a more careful 
analysis.  

As demonstrated in \cite{hep-th/0405001,hep-th/0412188,hep-th/0410105}, the detailed form of the
thermodynamic string Bethe equations prevents one from adopting the naive interpretation
of the function $\sigma(x)$, introduced in the spectral representation of the 
resolvent in eqn.~(\ref{spectralG}), as a density of string energy eigenstates 
supported on the contours $C_i$.  This problem can be summarized by noting that  
the first condition in eqn.~(\ref{RHBC})
\be
\int_C d{x}\, \sigma({x})  \sim \Delta+S-J
\ee
implies that the normalization of the spectral density is coupling-dependent, 
due to the presence of $\Delta$ on the right-hand side \cite{hep-th/0410105}.

In \cite{hep-th/0405001,hep-th/0410105}, it was shown that 
a legitimate excitation density $\rho$ (with coupling-independent normalization)
can be defined by introducing a nonlinear redefinition of the spectral parameter:
\be
\label{COV}
\varphi \equiv x + \frac{T}{x}\ ,
\ee
where $T \equiv \lambda' / 16\pi^2$, such that 
\be
\rho(\varphi) = \sigma(x)\ .
\ee
The quasi-momentum then depends on the spectral parameter $\varphi$ according to 
\be
\label{pCOV}
p(\varphi) = 1/\sqrt{\varphi^2 - 4 T}\ .
\ee  
Starting from the continuum Bethe equation in the deformed $\Sl(2)_\gamma$
sector of the string theory (eqn.~(\ref{RH2})), we therefore invoke the change of variables in
eqn.~(\ref{COV}) to obtain 
\be
\label{RH3}
2\pint d\varphi' \frac{\rho(\varphi')}{\varphi-\varphi'}
	& = & 2\pi (n_k - \Mgamma J) - p(\varphi)
\nn\\
&&\kern-50pt
	+ \pint d\varphi' \rho(\varphi')
	\biggl\{
	\frac{2T}{\sqrt{{\varphi'}^2-4T}\sqrt{{\varphi}^2-4T}}
	\left(\frac{x}{T-x x'} - \frac{x'}{T-x x'} \right)
\nn\\
&&\kern+30pt
	+ 4\pi \Mgamma J T 
	\left(
	\frac{1}{x^2-T} - \frac{1}{{x'}^2-T} 
	\right)
	\biggr\}\ .
\ee
To keep this equation compact, $x$ and $x'$ are understood via eqn.~(\ref{COV}) 
to be functions of $\varphi$ and $\varphi'$, respectively.

Following \cite{hep-th/0410105,hep-th/0405001}, we should be able to recast the integral term 
on the right-hand side of eqn.~(\ref{RH3}) strictly in terms of the 
relativistic dispersion relations
\be
\label{qrCONT}
q_r(\varphi) = \frac{1}{\sqrt{\varphi^2-4T}}
	\left(\frac{1}{2}\varphi+\frac{1}{2}\sqrt{\varphi^2-4T}\right)^{1-r}\ .
\ee
The $\Mgamma$-independent terms are known to arise from an infinite sum over these
$q_r(\varphi)$ \cite{hep-th/0406256}.  We find that the remaining terms coming from the deformation can be 
written simply as a linear expression in the dispersion relation $q_2(\varphi)$:
\be
\label{continuum}
2\pint d\varphi' \frac{\rho(\varphi')}{\varphi-\varphi'}
	& = & 2\pi (n_k - \Mgamma J) - p(\varphi)
\nn\\
&&\kern-60pt
	-2 \pint d\varphi' \rho(\varphi') \biggl\{
	\sum_{r=1}^\infty T^{r}\left(
	q_{r+1}(\varphi')q_{r}(\varphi)-q_{r}(\varphi')q_{r+1}(\varphi)
	\right)
	+2 \pi \Mgamma J  T \left(
	q_2(\varphi)- q_2(\varphi') \right)
	\biggr\}\ .
\nn\\
&&
\ee
Now, following \cite{hep-th/0406256,hep-th/0412188}, we interpret this continuum
equation as the thermodynamic limit of a discrete, or ``quantum'' Bethe
ansatz.  (To be sure, the thermodynamic limit is taken to render a distribution of Bethe roots
that is macroscopic and smooth in $\varphi$:  $J$ and $S$ become infinite, with the 
filling fraction $S/J$ held fixed.)  
One is then instructed to rely on the conjectured all-loop Bethe equations in the dual 
gauge theory to properly discretize this equation. 
The final quantum string Bethe equation in the deformed $\Sl(2)_\gamma$ sector thus
takes the form
\be
\label{FBE}
e^{i(p_k-2\pi \Mgamma ) J} = \prod_{j=1 \atop j \neq k}^S 
	\frac{\varphi(p_k) - \varphi(p_j) - i}{\varphi(p_k) - \varphi(p_j) + i}\,
	e^{- 2\pi i \Mgamma g^2  ( q_2(p_k)- q_2(p_j))}
	\prod_{r=1}^\infty e^{-2i\theta_r(p_k,p_j)}\ ,
\ee
where
\be
\theta_r(p_k,p_j) \equiv \left(\frac{g^2}{2}\right)^r \left(
	q_r(p_k) q_{r+1}(p_j) - q_{r+1}(p_k) q_{r}(p_j)
	\right)\ ,
\ee
and, as usual, $g^2 \equiv \lambda/8\pi^2$.
Remarkably, this equation precisely reproduces the $O(1/J)$ near-pp-wave
energy shift computed above in eqn.~(\ref{nearppSL2}). Furthermore, it is easy to
verify that the $\lambda^{1/4}$ strong-coupling behavior of the spectrum is not 
spoiled by the deformation.

The dispersionless scattering of excitations 
in the worldsheet theory is therefore conjectured to be encoded in the following 
two-body scattering phase:
\be
\label{phase}
\theta(p_k,p_j,\Mgamma) \approx -\frac{2}{\varphi(p_k)-\varphi(p_j)}
	-2 \sum_{r=1}^\infty \theta_r(p_k,p_j)
	- 2\pi \Mgamma g^2 
	\left(	q_2(p_k)- q_2(p_j) \right)\ .
\ee
The $\Mgamma$-dependent deformation term in $\theta(p_k,p_j,\Mgamma)$ is
determined by $q_2(p)$, which is the energy of a single excitation of momentum
$q_1(p) = p$.\footnote{Roughly speaking, this might be understood as arising from 
the fact that the TsT deformation considered here involves a shift in the timelike 
direction $\varphi_1$ \cite{Niklas/conversation}.}
To make contact with the thermodynamic limit, one invokes the
following rescaling:
\be
p_k \to p_k/J\ , \qquad q_r(p_k) \to J^{-r}q_r(p_k)\ .
\ee
It is then straightforward to see that eqn.~(\ref{continuum}) is properly embedded in
the discrete Bethe ansatz in eqn.~(\ref{FBE}).

The nontrivial accomplishment 
of the discretization is to capture the intricate dependence of the energy spectrum 
on the worldsheet momenta $p_k$, which is seen, for example, in eqn.~(\ref{nearppSL2}).
This information is washed out in the thermodynamic limit.  Furthermore, 
we point out that one does not need to ``discretize'' the relativistic formulas
for the rapidities $\varphi(p_k)$ or dispersion relations $q_r(p_k)$ from eqns.~(\ref{pCOV})
and (\ref{qrCONT}) to obtain the correct energy spectrum in the near-pp-wave 
limit.\footnote{We thank Juan
Maldacena for clarification on this point.}  
It has been suggested \cite{hep-th/0406256,hep-th/0412188}, however, 
that the proper ``lattice'' relations are indeed obtained by replacing the quantities 
$\varphi(p_k)$ and $q_r(p_k)$ with the following expressions, motivated by studies 
in the dual gauge theory:
\be
\varphi(p_k) &=& \frac{1}{2}\cot\left(\frac{p_k}{2}\right)\sqrt{1+8g^2\sin^2(p_k/2)}\ ,
\nn\\
q_r(p_k) &=& \frac{2\sin\left(\frac{r-1}{2}p_k\right)}{r-1}\left(
	\frac{\sqrt{1+8g^2\sin^2(p_k/2)}-1}{2g^2\sin(p_k/2)}
	\right)^{r-1}\ .
\ee
Rather remarkably, Hofman and Maldacena have recently made contact with these expressions
from the string side of the correspondence \cite{hep-th/0604135}.  (It should be noted
that \cite{hep-th/0509015} was an important precursor to this work;  
see also \cite{hep-th/0604175} for related developments).

Finally, we note that, with respect to strictly 
reproducing the energy shift in eqn.~(\ref{nearppSL2}), 
the two-body scattering phase in eqn.~(\ref{phase}) is not unique.  
Relying on near-pp-wave spectral information alone, it is straightforward to
formulate an ad hoc scattering phase that succeeds in reproducing 
eqn.~(\ref{nearppSL2}):
\be
\theta(p_k,p_j,\Mgamma) \approx -\frac{2}{\varphi(p_k)-\varphi(p_j)}
	-\left(2 + 4\pi\Mgamma\, \frac{p_k+p_j}{p_k\, p_j}\right)
	\sum_{r=1}^\infty \theta_r(p_k,p_j)\ .
\ee
This expression, however, is not connected in any transparent way to the 
thermodynamic Bethe ansatz in eqns.~(\ref{RH1},\ref{RH2}) above.

\section{Comparison with the gauge theory}
The construction of the Lunin-Maldacena solution has lead to a substantial 
amount of research on the dual $\beta$-deformed gauge theory (recent work includes 
\cite{hep-th/0505100,hep-th/0506128,hep-th/0506150,hep-th/0507113,hep-th/0507282,hep-th/0512194,hep-th/0602141}). 
The relationship of these theories to the spin-chain 
description was studied in \cite{hep-th/0312218,hep-th/0405215}, and
Bethe equations for the full twisted ${\cal N}=4$ SYM theory were constructed in
\cite{hep-th/0505187}. 
For the deformed ``$\su(2)_\beta$'' sector we need only consider a single type of impurity
in the (twisted) Bethe equations: 
\be
1={e}^{-2 \pi i \beta L}\left(\frac{u_k-i/2}{u_{k}+i/2}\right)^L\prod_{j\not=k}^{M}
\frac{u_k-u_j+i}{u_k-u_j-i}\ ,
\ee
where the cyclicity condition appears as
\be
1={e}^{2 \pi i \beta M}\prod_{j=1}^{M}\frac{u_k+i/2}{u_{k}-i/2}\ .
\ee
The corresponding equations to all loop-order in $\lambda$ 
are given by applying the same twist to the  
Bethe ansatz formulated by Beisert, Dippel and Staudacher (BDS) \cite{hep-th/0405001}
(these equations are still understood to be conjectural beyond three-loop order). 
It is then straightforward to compare $O(1/J)$ corrections 
to the string energy spectrum with corresponding predictions from the gauge theory
spin chain.  For the $\su(2)_\beta$ sector we find (as is now
expected) agreement up to three-loop order in $\lambda'$ and disagreement
at higher loops.  
As noted above, a Bethe ansatz for the twisted $\Sl(2)_\beta$ sector was also 
proposed in \cite{hep-th/0505187} by again taking a simple twist, analogous to
the $\su(2)_\beta$ sector.  This is indeed what we find from the string theory at one-loop order
in the $\Sl(2)_\gamma$ sector, but starting at two loops we find a more complicated 
dependence on the deformation parameter.

It was recently suggested \cite{hep-th/0512077} that the undeformed $\su(2)$ sector 
of the gauge theory might be described non-perturbatively by the Hubbard model.
(More precisely, the BDS Bethe equations arise in the weak-coupling limit of the Hubbard
model, so the correctness of this approach is tied to that of the BDS equations.)
Here we briefly describe how to modify this model to provide an analogous
description in the $\beta$-deformed gauge theory.\footnote{Following the completion of this work, we
were notified that similar results are derived in \cite{Ramos:1996us,martins-1997-}.} 
Starting from the Hubbard model Hamiltonian
\be
H_{{\rm{Hubbard}}}=-t \sum_{n,\ \sigma=\uparrow, \downarrow}^L 
(c^{\dagger}_{n,\sigma}c_{n+1,\sigma}+ c^{\dagger}_{n+1,\sigma}c_{n,\sigma})
+t U \sum_{n=1}^L c^{\dagger}_{n,\uparrow}
c_{n,\uparrow}c^{\dagger}_{n,\downarrow}c_{n,\downarrow}\ ,
\ee
we wish to study the following generalization:
\be
	H_{\rm{Deformed}}=-t  \sum_{n,\ \sigma=\uparrow, \downarrow}^L( f(n,\sigma)
	c^{\dagger}_{n,\sigma}c_{n+1,\sigma}+ {\tilde f}(n,\sigma)c^{\dagger}_{n+1,\sigma}
	c_{n,\sigma})+t U \sum_{n=1}^L c^{\dagger}_{n,\uparrow}c_{n,\uparrow}
	c^{\dagger}_{n,\downarrow}c_{n,\downarrow}\ .
\ee
Now, exactly as in \cite{hep-th/0512077}, one may calculate the effective action 
in this twisted model to first order in perturbation theory. 
Briefly, one takes the following Hamiltonian at leading order
\be 
H_0=t U \sum_{n=1}^L c^{\dagger}_{n,\uparrow}c_{n,\uparrow}
	c^{\dagger}_{n,\downarrow}c_{n,\downarrow}\ , 
\ee
and considers a subspace of the complete Fock space spanned by the states 
$c^{\dagger}_{1\sigma_1}c^{\dagger}_{2\sigma_2}\ldots c^{\dagger}_{L\sigma_L}|0\rangle$. 
It is then straightforward to show that the effective 
one-loop Hamiltonian in this particular subspace is 
\be
h&=&-\frac{1}{2} \sum_n 2\left(f(n,\uparrow){\tilde f}(n,\uparrow)+f(n,\downarrow)
{\tilde f}(n,\downarrow)\right)\left((S_n^z S_{n+1})^z-\frac{1}{4}\right)\nn\\
& &+\left(f(n,\uparrow){\tilde f}(n,\uparrow)+f(n,\downarrow){\tilde f}(n,\downarrow)
\right)\left(S^z_{n+1}-S^z_{n}\right)\nn\\
& &+2 f(n,\uparrow){\tilde f}(n, \downarrow)S^+_n S^-_{n+1}+2 f(n,\downarrow)
{\tilde f}(n, \uparrow)S^-_n S^+_{n+1}\ .
\ee
We have used 
\be
  S^+_n=c_{n,\uparrow}^{\dagger}c_{n,\downarrow}\ , \qquad S^-_n=c_{n,\downarrow}^{\dagger}
       c_{n,\uparrow}\ ,
\ee
along with the following: 
\be
  S^z_z=\frac{1}{2}\left(c_{n,\uparrow}^{\dagger}c_{n,\uparrow}-c_{n,\downarrow}^{\dagger}
     c_{n,\downarrow}\right)
	\simeq
     c_{n,\uparrow}^{\dagger}c_{n,\uparrow}-\frac{1}{2}
	\simeq
	\frac{1}{2}-c_{n,\downarrow}^{\dagger}
     c_{n,\downarrow}\ ,
\ee
which only hold true when acting on singly-occupied states. 
One may then compare this Hamiltonian with the one-loop 
$\beta$-deformed $\su(2)_\beta$ spin-chain Hamiltonian, 
formulated in \cite{hep-th/0503192}:
\be
\label{betaspin}
H&=&	\frac{|h|^2}{2}\sum_{n=1}^L\Bigl(\cosh 2\pi\kappa_d\left(S^z_n S^z_{n+1}-1/4\right)
		+1/2\sinh 2\pi\kappa_d \left(S^z_n-S_{n+1}^z\right)
\nn\\
& & 
\kern+45pt
		+\frac{1}{2}e^{2 \pi i \beta}S^+_nS^-_{n+1}+\frac{1}{2}e^{-2 \pi i \beta}
		S^-_{n}S^+_{n+1}\Bigr)\ .
\ee
This form is slightly more general than we require for comparison with 
the $\gamma$-deformed string theory, for which we can set 
$|h|=1$ and $\kappa_d=0$. By matching coefficients
\be
f(m,\uparrow)=|h|e^{i \pi(\beta+\kappa_d)}\ ,&\qquad& {\tilde f}(m,\downarrow)
=|h|e^{i \pi(\beta-\kappa_d)}\ ,\nn\\
f(m,\downarrow)=|h|e^{-i \pi(\beta+\kappa_d)}\ ,&\qquad& {\tilde f}(m,\uparrow)
=|h|e^{-i \pi(\beta-\kappa_d)}\ ,
\ee
we precisely reproduce the $\beta$-deformed $\su(2)_\beta$ spin chain Hamiltonian in
eqn.~(\ref{betaspin}).

We note that the $\beta$-deformed Hamiltonian with $|h|=1$ can be reached from the undeformed
model with twisted boundary conditions under
\be
c_{m,\uparrow}& \rightarrow &c_{m, \uparrow}e^{i \pi \beta m}e^{\pi \kappa_d/2}\ ,\nn\\
c_{m,\downarrow}&\rightarrow &c_{m, \downarrow}e^{-i \pi \beta m}e^{-\pi \kappa_d/2}\ .
\ee
The matrix generating this transformation is
\be
U=\begin{pmatrix}e^{i \pi \beta m}e^{\pi \kappa_d/2}&&0\\0&&e^{-i \pi \beta m}
e^{-\pi \kappa_d/2}\end{pmatrix}\ ,
\ee
or, equivalently,
\be
U= \exp \left(\frac{i \pi}{2}\sum_m (c_{m,\uparrow}^{\dagger}c_{m,\uparrow}
-c_{m,\downarrow}^{\dagger}c_{m,\downarrow})(\beta m-i \kappa_d/2)\right)\ .
\ee
In the case with $|h|=1$, $\kappa_d=0$, this transformation is unitary and  
the deformed Hubbard model merely corresponds to imposing 
twisted boundary conditions on the corresponding spin chain, with different conditions 
for the spin-up and spin-down fermions.

The Bethe equations solving this deformed 
Hubbard model arise as simple twists of the undeformed Lieb-Wu 
\cite{1968PhRvL..20.1445L,1968PhRvL..21..192L,lieb-2003-321} equations.  
Twisted Lieb-Wu equations of this sort were studied by Yue and Deguchi 
in \cite{cond-mat/9606039} (see also appendix C of \cite{hep-th/0512077}, to which 
we also refer for notation):  
\be
\label{hcycl}
&&  {e}^{i{\tilde q}_n L}=
	\prod_{j=1}^M\frac{u_j-\sqrt{2}g\sin({\tilde q}_n+\phi_{\uparrow})-i/2}
	{u_j-\sqrt{2}g\sin({\tilde q}_n+\phi_{\uparrow})+i/2}\ , \qquad  n=1,\ldots,N
\nn\\
\label{hbethe}
&& 	\prod_{k=1}^N\frac{u_k-\sqrt{2}g\sin({\tilde q}_n+\phi_{\uparrow}+i/2}
	{u_k-\sqrt{2}g\sin({\tilde q}_n+\phi_{\uparrow})-i/2}
	={e}^{iL(\phi_{\downarrow}-\phi_{\uparrow})}
	\prod_{j\not= k}^M
	\frac{u_k-u_j+i}{u_k-u_j-i}\ .
\ee
The energy eigenvalues are computed by solving the Bethe equations for the momenta
$\tilde q_n$ and using the formula
\be
\label{henergy}
E=\frac{\sqrt{2}}{g}\sum_{n=1}^{L}\cos({\tilde q}_n+\phi_{\uparrow})\ .
\ee
As in \cite{hep-th/0512077}, we have chosen Hubbard couplings that make the connection
to the gauge theory obvious. 
For our twisted Hubbard model we set $\phi_{\uparrow}=\pi \beta$ and 
$\phi_{\downarrow}=-\pi \beta$, and it is 
straightforward to see that at half-filling (i.e., when the number of 
fermions $N$ equals the lattice 
length $L$) and in the weak coupling limit ($g\rightarrow 0$), 
the second equation in (\ref{hbethe}) reduces to the 
one-loop twisted Bethe equations for the gauge theory in the $\su(2)_\beta$ sector.

To see that the energy spectrum behaves as expected under this twist, 
it is useful to perform the transformation introduced in \cite{hep-th/0512077}, 
similar to a Shiba transformation.  With the definitions
\be
c_{n,\circ}=c^{\dagger}_{n,\uparrow}\ ,\qquad c_{n,\circ}^{\dagger}=c_{n,\uparrow}\ ,\nn\\
c_{n,\updownarrow}=c_{n,\downarrow}\ ,\qquad c_{n,\updownarrow}^{\dagger}=c_{n,\downarrow}^{\dagger}\ ,
\ee
we rewrite the twisted Hamiltonian in its dual form
\be
H&=&\frac{1}{\sqrt{2}g}\sum_{n=1,\ \sigma=\circ,\updownarrow}^{L}
\left({e}^{i \phi_{\sigma}} 
c_{n,\sigma}^{\dagger}c_{n+1,\sigma}+{e}^{-i \phi_{\sigma}}
c_{n+1,\sigma}^{\dagger}c_{n,\sigma}\right)
\nn\\
& &
\kern+10pt
	-\frac{1}{g^2}\sum_{n=1}^{L}\left(1-c_{n,\circ}^{\dagger}
	c_{n,\circ}\right)c_{n,\updownarrow}^{\dagger}
	c_{n,\updownarrow}\ ,
\ee
where $\phi_{\updownarrow}=\phi-\pi \beta,\ \phi_{\circ}=\pi-(\phi+\pi \beta)$. 
The parameter $\phi$ is analogous to Aharonov-Bohm flux \cite{hep-th/0512077}:
it is chosen to be $\phi=0$ for $L={\rm odd}$ and $\phi={\pi}/{2 L}$ for $L={\rm even}$. 
The Bethe equations for the dual Hamiltonian at half filling take the form
\be
\label{hdcycl}
&&
	e^{iL ({\tilde q}_n + \pi \beta)}=
	\prod_{j=1}^M\frac{u_j-\sqrt{2}g\sin({\tilde q}_n-\phi)-i/2}
	{u_j-\sqrt{2}g\sin({\tilde q}_n-\phi)+i/2}\ ,\quad  n=1,\ldots,2M\ ,
\\
\label{hdbethe}
&&	
	\prod_{n=1}^{2 M}\frac{u_k-\sqrt{2}g\sin({\tilde q}_n-\phi)+i/2}
	{u_k-\sqrt{2}g\sin({\tilde q}_n-\phi)-i/2}=-
	\prod_{j\not= k}^M
	\frac{u_k-u_j+i}{u_k-u_j-i}\ ,
\ee
where the energy is now given by 
\be
\label{hdenergy}
E=-\frac{M}{g^2} - \frac{\sqrt{2}}{g}\sum_{n=1}^{2 M}\cos({\tilde q}_n-\phi)\ .
\ee

It is straightforward to study the effect of the $\beta$ deformation on this Hubbard
model:  one route is to solve the one-magnon problem, as in \cite{hep-th/0512077},
with a system composed of $M=1$ down spins and $L-1$ up spins.  Following
\cite{hep-th/0512077}, we adopt the ansatz
\be
\tilde q_1 -\phi = \frac{\pi}{2} + q + i\delta\ , \quad 
\tilde q_2 -\phi = \frac{\pi}{2} + q - i\delta\ ,
\ee
where $\delta$ parameterizes the binding of the quasi-momenta $\tilde q_n$.
We may then use the set of one-magnon Bethe equations
\be
\label{hdbethe1}
&&	e^{i L (\tilde q_1 + \pi \beta)}=\frac{u-\sqrt{2}g\sin(\tilde q_1-\phi)-i/2}{u-\sqrt{2}g\sin(\tilde q_1-\phi)+i/2}\ ,
\quad
	e^{i L(\tilde q_2 + \pi \beta)}=\frac{u-\sqrt{2}g\sin(\tilde q_2-\phi)-i/2}{u-\sqrt{2}g\sin(\tilde q_2-\phi)+i/2}\ ,
\nn\\
&&
\ee
and
\be
	\frac{u-\sqrt{2}g\sin(\tilde q_1-\phi)+i/2}{u-\sqrt{2}g\sin(\tilde q_1-\phi)-i/2}
	\frac{u-\sqrt{2}g\sin(\tilde q_2-\phi)+i/2}{u-\sqrt{2}g\sin(\tilde q_2-\phi)-i/2}=-1\ ,
\ee
to find
\be
q = \frac{\pi}{L}(n - \beta L)\ , \qquad n = 0,1,\ldots,L-1\ .
\ee
The magnon momentum is defined to be $p \equiv 2 q$, so we see that the twist indeed 
amounts to a shift in the mode number $n$ by $\beta L$.  We can now show that the energy $E$, written as a dispersion law in $p$, yields the form we expect in the
presence of the twist (this turns out to be a nontrivial issue).  
Rewriting the equations in eqn.~(\ref{hdbethe1}) as 
\be
\label{Bdrop}
\sqrt{2}g\,\sin(\tilde q_{1,2}-\phi)-u = \frac{1}{2}\cot \left(\frac{(\tilde q_{1,2}+\pi \beta)L}{2}\right)\ ,
\ee
one finds that, expressed in terms of mode numbers, 
all instances of $\beta$ drop out of this equation.
By splitting into real and imaginary pieces, we therefore obtain
\be
&& \sinh(\delta)=\frac{\tanh(\delta L)}{2\sqrt{2} g \sin(q)}\ , 
\nn\\
&& u=\sqrt{2}g \cos(q)\cosh(\delta)+\frac{(-1)^n(-1)^{\frac{L+1}{2}}}{2\cosh(\delta L)}\ .
\ee
In the limit $L \to \infty$, terms of the form $e^{-\delta L}$ are dropped, and, following \cite{hep-th/0512077},
we find that the energy formula for the one-magnon system is
\be
E=-\frac{1}{g^2}-\frac{2 \sqrt{2}}{g}\sin(p/2)\cosh(\delta)=
	-\frac{1}{g^2}+\frac{1}{g^2}\sqrt{1+8g^2\sin^2 \left(\frac{\pi}{L}(n-\beta L)\right)}\ .
\ee

The full $M$-magnon problem can be solved in a 
completely analogous fashion, in which case the quasi-momenta $\tilde q_n$ are split into two sets:
\be
\tilde q_n - \phi & = & s_n \frac{\pi}{2} + \frac{p_n}{2} + i\delta_n\ , \nn\\
\tilde q_{n+M} - \phi & = & s_n \frac{\pi}{2} + \frac{p_n}{2} - i\delta_n\ ,
\ee
where $s_n = {\rm sign}(p_n)$, and $n = 1,\ldots,M$. Once again, we take $L\to \infty$ and obtain 
\be
e^{i \tilde q_n L}&\sim &e^{-\delta_n L}\to 0\ , \nn\\
e^{i \tilde q_{n+M} L}&\sim &e^{\delta_{n+M} L}\to \infty\ .
\ee
Thus, for $L$ large, there exists a $u$ denoted by $u_n$ for each $n\in 1,\ldots, M$, such that 
\be
u_n-i/2=\sqrt{2} g \sin (\tilde q_n -\phi)\ , \qquad  u_n+i/2=\sqrt{2} g \sin (\tilde q_{n+M} -\phi)\ ,
\ee
exactly as in \cite{hep-th/0512077}. We can use these equations to to determine $\delta_n$ and $u_n$ in terms of $p_n$, 
and we find the same expressions as in the undeformed case:
\be
\sinh \delta_n&=&\frac{1}{2 \sqrt{2} g |\sin \frac{p_n}{2}|}\ ,
\\
u_n&=& \frac{1}{2} \cot \frac{p_n}{2}\sqrt{1+8 g^2 \sin^2\frac{p_n}{2}}\ .
\ee 
To determine $\tilde{q}_{n}$ in terms of $u_n$, we multiply the $n^{\rm th}$ and $(n+M)^{\rm th}$ 
equation in (\ref{hdbethe}) and it is straightforward to see that 
\be
e^{i L\left(p_n+2 \pi \beta\right)}=\prod^M_{j=1,\ j\not= n}\frac{u_n-u_j+i}{u_n-u_j-i}\ .
\ee
We are therefore lead to conclude that the twisted BDS Bethe equations
are properly encoded in this $\beta$-deformed Hubbard model, which, given that the deformation is
merely a twisted boundary condition, is to be expected.

\section{Conclusions}
TsT transformations yield a simple deformation
of the usual correspondence between string theory in $AdS_5\times S^5$ and ${\cal N}=4$
SYM theory.   Many of the recent developments stemming from the discovery of integrable
structures in this correspondence are thus easily tested in this interesting new setting.  
In recent years, for example, 
a heuristic methodology has emerged for formulating quantum string Bethe 
equations from Lax representations of string sigma models.  Thermodynamic
Bethe equations emerge directly from the sigma model
by formulating the Bethe ansatz as a Riemann-Hilbert problem.  However, one must
rely on detailed studies of the dual gauge theory for instruction on how to
discretize these equations.  In this paper we have studied   
whether the discretization procedure handed down from the gauge theory 
can be applied in the case of TsT-deformed string theory.   We have shown that these rules 
can indeed be adopted under relatively dramatic deformations of the original problem, and we 
have been able to successfully reproduce $O(1/J)$ 
corrections to the plane-wave energy spectrum in deformed $\Sl(2)_\gamma$ and 
$\su(2)_\gamma$ subsectors.  

We have also made contact in the $\su(2)_\gamma$ sector with a recent formulation
of the all-loop gauge theory problem written as a low-energy effective theory
embedded in the Hubbard model.  
One open problem is that it is difficult to make any such contact with the gauge theory 
side of the correspondence in the $\Sl(2)_\gamma$ sector.  
General considerations lead us to believe that the field theory dual to this deformed
string theory is a non-commutative gauge theory.  
It would be very interesting to study whether a non-commutative deformation of
${\cal N}=4$ SYM theory encodes 
some portion of the $\Sl(2)_\gamma$ string spectrum computed here.   

Furthermore, one can extend the analysis of the near-pp-wave theory to larger subsectors
of excitations. At one-loop order in $\lambda'$ one might expect the full set of $\so(6)$
bosons to comprise a closed subsector, sensitive to all three deformation 
parameters of the non-supersymmetric $\gamma$-deformed background.  
Even at one-loop order, however, string
theory predictions disagree with corresponding anomalous dimensions in the gauge theory. 
It is possible that in the non-supersymmetric deformation the $\so(6)$ bosons do not form 
a closed subsector but mix with the fermions.  Another possibility is that 
the background itself needs to be corrected, perhaps along the lines of \cite{hep-th/0603207}. 
It is also interesting that the 
non-supersymmetric deformation is unstable due to the flow of double trace couplings 
\cite{Roiban/conversation}, similar to the case found for orbifolds of 
${\cal N}=4$ SYM \cite{hep-th/0505099,hep-th/0509132}. 
For the deformed theory, however, the endpoint of the flow is unclear. 

At this point there exist a number of quantum string Bethe equations 
harboring a great deal of predictive information that remains untested.
In this regard it would obviously be valuable to obtain spectral information 
directly from the string theory at higher orders in the $1/J$ expansion.  Attempts 
to study this difficult problem seem to be hindered by the lack of a suitable 
renormalization scheme for the Green-Schwarz formulation of the worldsheet 
lightcone field theory. Perhaps a covariant approach is needed to glean reliable 
information beyond the near-pp-wave limit.
Alternatively, it would be extremely valuable to uniquely derive the complete $S$ matrix
of the worldsheet theory based on the underlying symmetries in the problem.

\section*{Acknowledgments}
We thank Niklas Beisert, Juan Maldacena, Andy Neitzke, Radu Roiban, 
Marcus Spradlin and Anastasia Volovich for 
interesting discussions and useful comments.  TMcL is supported by Pennsylvania State University.
IS is supported as the Marvin L.~Goldberger member at the Institute for Advanced Study, 
and by National Science Foundation grant PHY-0503584. 

\bibliographystyle{utcaps}
\bibliography{Deformed_AdSCFT_v11}

\end{document}